\documentclass[pre, superscriptaddress, secnumarabic, twocolumn,showpacs,showkeys,amsmath,amssymb,nofootinbib,tightenlines]{revtex4}

\usepackage{graphicx}% Include figure files

\newcommand{\vect}[1]{\boldsymbol {#1}}

\newcommand{\ti}[1]{#1}  %keep titles, etc, of refs
\newcommand{\vol}[1]{\bf #1}  %journal volume style
\newcommand{\E}{\Upsilon} %{\mathcal{E}}
\newcommand{\Zup}{\v{Z}upanovi\'c }
\newcommand{\alf}{a}

\DeclareMathOperator{\sign}{sign}

\allowdisplaybreaks[4]
\begin{document}

\title{Simultaneous Extrema in the Entropy Production \\ for Steady-State Fluid Flow in Parallel Pipes}% Force line breaks
\author{Robert K. Niven}
\email{r.niven@adfa.edu.au}
\affiliation{School of Engineering and Information Technology, The University of New South Wales at ADFA, Canberra, ACT, 2600, Australia.}
%\email{r.niven@adfa.edu.au}
%end aps

\date{26 November 2009; {revised 11 February 2010}}% can use \today

\begin{abstract}

Steady-state flow of an incompressible fluid in parallel pipes can simultaneously satisfy two contradictory extremum principles in the entropy production, depending on the flow conditions. For a constant total flow rate, the flow can satisfy (i) a pipe network minimum entropy production (MinEP) principle with respect to the flow rates, and (ii) the maximum entropy production (MaxEP) principle of {Ziegler and Paltridge} with respect to the choice of flow regime.  The first principle - different to but allied to that of Prigogine - arises from the stability of the steady state compared to non-steady-state flows; it is proven for isothermal laminar and turbulent flows in parallel pipes with a constant power law exponent, but is otherwise invalid.  The second principle appears to be more fundamental, driving the formation of turbulent flow in single and parallel pipes at higher Reynolds numbers.  For constant head conditions, the flow can satisfy (i) a modified maximum entropy production (MaxEPMod) principle of \Zup and co-workers with respect to the flow rates, and (ii) an inversion of the {Ziegler-Paltridge} MaxEP principle with respect to the flow regime. The interplay between these principles is demonstrated by examples.
 
\end{abstract}

%aps format

\maketitle
%end aps
%
%$\mathcal{abcdefghijklmnopqrstuvwxyz}$\\
%$\mathcal{ABCDEFGHIJKLMNOPQRSTUVWXYZ}$\\
%$\mathscr{abcdefghijklmnopqrstuvwxyz}$\\
%$\mathscr{ABCDEFGHIJKLMNOPQRSTUVWXYZ}$\\
%$\mathfrak{abcdefghijklmnopqrstuvwxyz}$\\
%$\mathfrak{ABCDEFGHIJKLMNOPQRSTUVWXYZ}$\\
%

%% %%############################################################################
\section{\label{Intro}Introduction} 
%% ############################################################################
For millennia, the discipline of {\it fluid mechanics} has inspired many of humanity's deepest thinkers, including Archimedes, Al-Khazini, da Vinci, Torricelli, Newton, Pascal, the Bernoullis, Euler, d'Alembert, Lagrange, Laplace, Navier, Stokes, Helmholtz, Reynolds, Taylor, Rayleigh, Prandtl, Onsager and Kolmogorov.  Despite this long and rich history, and the more recent development of thermodynamics over the past 150 years, surprisingly few workers have sought to establish the connection between fluid mechanics and what is generally referred to as {\it non-equilibrium thermodynamics}: the study of systems which experience one or more flows of various physical quantities in response to thermodynamic gradients.  Of particular importance to the latter is the concept of {\it entropy production}, as defined by \cite{deGroot_M_1962, Prigogine_1967, Kondepudi_P_1998}:
\begin{equation}
\dot{\sigma} = \sum\limits_j \vect{F}_j \cdot \vect{j}_j
\label{eq:EP}
\end{equation}
where $\vect{F}_j$ is the $j$th thermodynamic force acting on the system and $\vect{j}_j$ is the corresponding flux of the $j$th physical quantity. Examples of the forces include the gradients $\nabla (1/T)$, %$-\nabla (p/T)$, 
$-\nabla (\mu_c/T)$, $-\nabla (\vect{v}/T)$ and $\vect{E}/T$, where $k$ is Boltzmann's constant, $T$ is absolute temperature, %$p$ is absolute pressure, 
$\mu_c$ is the mass chemical potential of the $c$th constituent, $\vect{v}$ is the mass-average fluid velocity and $\vect{E}$ is the electric field. These generate, respectively, corresponding fluxes of heat $\vect{j}_q$, %fluid mass $\vect{j}_f$, 
mass of each constituent $\vect{j}_c$, momentum $\vect{\tau}$ and charge $\vect{j}_z$ \cite{deGroot_M_1962, Prigogine_1967, Kondepudi_P_1998}. The lack of association with fluid mechanics is even more surprising with the knowledge that non-equilibrium thermodynamics is essentially the science of {\it dissipative systems} \cite{Prigogine_1967, Prigogine_1980, Prigogine_S_1984, Kondepudi_P_1998} -- those which dissipate available energy as heat -- of which frictional fluid flow is a prominent example. Remarkably few reference books in fluid mechanics even mention the entropy production, whilst many of those that do \citep[e.g.][]{Bird_etal_2006, Munson_etal_2009} examine it in passing as merely another transport-related quantity, seemingly of no great importance. The connections between entropy production and the dissipation of available energy (exergy), heat generation, work expenditure and power consumption are usually not made explicit. Part of the problem is that fluid mechanics now tends to reside between the disciplines of physics and engineering; today, few physics degree programs contain more than a single course in fluid mechanics (if any!), whilst few engineering programs devote adequate time to ``esoteric'' concepts such as entropy.  Of course, exceptions to this lack of association do exist. Most relevant to the present analysis include several detailed reference works by Bejan \cite{Bejan_1982, Bejan_1996, Bejan_2006}, written for a somewhat different purpose (see \S\ref{EP_principles}), the quite farsighted work of Paulus Jr and co-authors \cite{PaulusJr_2000, PaulusJr_G_2004} (which, however, contains some errors), and a recent note by Martyushev \cite{Martyushev_2007}.

In thermodynamics, undoubtedly the most far-reaching advance was the development of {extremum} principles to determine the equilibrium position of a system. These include maximisation of the thermodynamic entropy of a closed system \cite{Clausius_1865, Boltzmann_1877} and its dual principle, minimisation of the free energy of an open system \cite{Gibbs_1875, Callen_1985}. In non-equilibrium thermodynamics, several extremum methods have been proposed to determine the stationary (steady) state \cite{Martyushev_S_2006, Martyushev_etal_2007, Bruers_2007c}. The first was advocated by Onsager \cite{Onsager_1931a, Onsager_1931b, Onsager_M_1953}, involving maximisation of the entropy production less a dissipative term, widely but inaccurately referred to as a ``minimum dissipation'' method. This method is connected to other modified maximum entropy production (MaxEPMod) principles for the analysis of viscous fluid flows \cite{Helmholtz_1868, Rayleigh_1913, Horne_K_1988} and electrical circuits \cite{Zupanovic_etal_2004, Botric_etal_2005, Christen_2006, Bruers_etal_2007a}. The next (and best-known) non-equilibrium extremum method involves minimisation of the entropy production (MinEP), expounded over many years by Prigogine \cite{Prigogine_1967, Prigogine_1980, Prigogine_S_1984, Kondepudi_P_1998}. This method has inspired or is connected to other MinEP (or minimum heat production) methods for the analysis of electrical \cite{Jeans_1966, Landauer_1975, Jaynes_1980, Bruers_etal_2007a, Kondepudi_P_1998} and fluid flow systems \cite{PaulusJr_2000, PaulusJr_G_2004}.  A third, more elusive principle has been proposed in several fields (e.g.\ by {Ziegler and Paltridge}) based on empirical evidence, involving maximisation of the entropy production (MaxEP) \cite{Paltridge_1975, Paltridge_1978, Ziegler_1977}.  This has been used to predict the steady-state behaviour of a variety of non-equilibrium ``complex systems'', including heat transport by convection in heated vessels (B\'enard cells) \cite{Ozawa_etal_2001}, planetary atmospheres \cite{Paltridge_1975, Paltridge_1978, Ozawa_etal_2001, Kleidon_etal_2003, Ozawa_etal_2003, Kleidon_L_book_2005} and the mantle \cite{Vanyo_Paltridge_1981}; global water and nutrient cycles \cite{Kleidon_2004, Kleidon_L_book_2005, Kleidon_S_2008}; crystal growth and diffusion \cite{Martyushev_A_2003, Christen_2007b}; resource degradation in ecosystems \cite{Meysman_B_2007, Bruers_M_2007}; biochemical reaction mechanisms \cite{Juretic_Z_2003, Dewar_etal_2006} and earthquake frequencies \cite{Main_Naylor_2008}. Other extremum principles have also been given for other purposes, e.g.\ methods based directly on maximising the Shannon entropy (MaxEnt) \cite{Awumah_etal_1990, Tanyimboh_T_1993, deSchaetzen_etal_2000, Formiga_etal_2003, Ang_J_2003, Setiadi_etal_2005}.

The aim of this work is to examine various extremum principles for the analysis of fluid flows in pipes, both individually and in parallel. The more difficult problem of flow in a generalised flow network -- in which not all nodes need be connected -- is deferred to a later analysis. Firstly, the various formulations of these principles are classified (\S\ref{EP_principles}), and the principles of fluid flow in a pipe are reviewed in light of their connection to frictional losses and hence entropy production (\S\ref{pipeflow}).  This leads into separate analyses of entropy production phenomena in single (\S\ref{single}) and parallel pipes (\S\ref{parallel}), the latter under both constant total flow rate or constant head conditions. In a single pipe, empirical data provide strong evidence that the {Ziegler-Paltridge} MaxEP principle applies, driving the formation of turbulent flow at higher Reynolds numbers. In parallel pipes, two simultaneous extremum principles can apply, depending on the flow conditions. The interplay of these effects is demonstrated using example systems.

%############################################################################
\section{\label{EP_principles}Entropy Production Extremum Principles} 
%% ############################################################################
%
Owing to considerable confusion in the literature, it is important to clarify the various entropy production extremum principles and their applicability to fluid flow in single and parallel pipes.  Firstly, just as the purpose of extremum principles in thermodynamics is to predict the equilibrium position, as a function of the intensive and/or extensive variable constraints \cite{Callen_1985}, we should expect that any extremum principle of interest here will enable prediction of the steady state of a flow system, as a function of constraints on the fluxes and/or thermodynamic forces. Extremum methods to obtain non-steady-state flows are therefore not examined. Further discussion of this point, in connection with a classification of different types of systems, is given elsewhere \cite{Niven_MEP, Niven_MaxEnt09, Niven_PhilTransB}. Secondly, the entropy production in a pipe flow (or electrical) network -- although calculated using \eqref{eq:EP} -- is of quite different character to that implied by \eqref{eq:EP}, since the index $j$ now refers to each conduit rather than each phenomenon \cite{Zupanovic_etal_2004, Martyushev_etal_2007}; typically, only one phenomenon -- such as fluid flow -- is of interest. In consequence, extremum principles designed for multi-phenomenon systems need not apply to pipe flow networks (and vice versa). Failure to recognise this distinction has caused a great deal of confusion in the literature. 

{It is of course recognised that the term ``steady state'' is something of a misnomer, since it refers only to a constant mean flow and not its temporal or spatial variability. Indeed, it has been shown that dissipative flow systems, such as fluid turbulence and heat convection, are {\it required} to exhibit variability at steady state \cite{Niven_MEP, Niven_PhilTransB}. For consistency with other studies, ``steady state'' is here used to describe a stationary state of constant mean flow, to distinguish it from thermodynamic equilibrium, but this does not imply lack of variability of the flow.}

Examining the available extremum principles in turn:
\begin{list}{$\bullet$}{\topsep 0pt \itemsep 3pt \parsep 0pt \leftmargin 8pt \rightmargin 0pt \listparindent 0pt
\itemindent 0pt}
\item Onsager's principle \cite{Onsager_M_1953} applies to flows in the linear transport regime, i.e.\ those in which the fluxes are linear functions of the forces $\vect{j}_j = \sum\nolimits_k L_{jk} \vect{F}_k$, where $j$ and $k$ are phenomenon indices and $L_{jk}$ are the (constant) phenomenological coefficients \cite{Onsager_1931a, Onsager_1931b}. The principle involves maximisation of $\dot{\sigma}$ less the dissipation function $\Phi=\frac{1}{2} \sum\nolimits_{jk} L_{jk}^{-1} \vect{j}_j \vect{j}_k$, where $L_{jk}^{-1}$ is the inverse matrix of $L_{jk}$.  The basis of the principle is obscure, but stems from  Markovian decay of a system towards equilibrium \cite{Onsager_M_1953}.  Since it requires multiple phenomena, this principle is not considered further here.

\item Prigogine's MinEP principle \cite{Prigogine_1967, Prigogine_1980, Prigogine_S_1984, Kondepudi_P_1998}, again applicable only in the linear transport regime, involves minimisation of $\dot{\sigma}$ with respect to certain fluxes $\vect{j}_k$, subject to a partial set of fixed forces $\vect{F}_k$. The theorem is readily proved in the assumption of local equilibrium.  This principle also requires multiple phenomena, and so is not considered further here.

\item The MaxEP principle of {Ziegler and Paltridge} is of quite different character, being advocated for systems well beyond the linear regime, subject to one or more fixed forces and/or fluxes. Due to the formation of non-linear response mechanisms, such systems can respond in many ways to the driving forces and fluxes, or in other words, exhibit multiple (stable) steady states. The MaxEP principle holds that the observed steady state is that which maximises the entropy production of the system. Aside from empirical evidence in many different fields (\S\ref{Intro}), several attempts have been made to derive this principle, including path-based analyses of transient non-equilibrium systems \cite{Dewar_2003, Dewar_2005, Attard_2006a, Attard_2006b}, a fluctuation analysis \cite{Zupanovic_etal_2006} and a relative velocity argument \cite{Martyushev_2007b}. Recently, the author \cite{Niven_MEP, Niven_MaxEnt09, Niven_PhilTransB} has given a local, conditional derivation of this principle by a probabilistic analysis of instantaneous fluxes through each element of a flow system. The analysis is identical to that used in thermodynamics to infer the equilibrium position, but adopts new entropy and free-energy-like functions defined for a flow system, so as to infer the steady-state position. Since the {Ziegler-Paltridge} MaxEP principle is considered to apply to single and multiple phenomena, it is examined here.

\item A separate MinEP principle has been developed in engineering circles, especially by Bejan \cite{Bejan_1982, Bejan_1996, Bejan_2006}, in which the entropy production of an engineered system is minimised to determine the most efficient design, {e.g.\ that with the highest energy efficiency}.  Accordingly, this is a {\it design} principle, which serves a different purpose to the {\it predictive} principles of interest here.  

\item {Related to this discussion is an observed feature of many natural and engineered flow systems, described by Bejan as the ``constructal law'', in his words \cite{Bejan_2006, Bejan_Lorente_2006, Bejan_2007}:
%\begin{quote}
``{\it For a finite-size flow system to persist in time ... it must evolve in such a way that it provides easier and easier access to the currents that flow through it.}''
%\end{quote}
This ``law'' appears to be connected in some manner to an entropy production extremum principle, since the system minimises its resistance for a given flow rate and/or maximises its flow rate for a given resistance. Although important, evolvable flow systems are not examined further in this study.}

\item A separate class of MinEP principles has been developed in finite-time thermodynamics, to calculate the fundamental limit of efficiency of a process in which a thermodynamic system is moved from one equilibrium position to another, along a specified path at specified rates \cite{Salamon_A_G_B_1980, Salamon_B_1983, Nulton_etal_1985, Andresen_G_1994, Crooks_2007}. The geodesic is used to determine the most efficient path. This method has been generalised to steady-state flow systems \cite{Niven_A_2009}. Of course, it is always possible to do worse (produce more entropy) than these limits. Since this method relates to transitions between equilibrium or steady states, not the prediction of the states themselves, it is not examined further here.

\item A separate MinEP (or minimum heat production) principle has been advocated for the analysis of electrical circuits \cite{Jeans_1966, Landauer_1975, Jaynes_1980, Kondepudi_P_1998} and more recently for fluid flow networks \cite{PaulusJr_2000, PaulusJr_G_2004}. This involves minimising \eqref{eq:EP} -- with $j$ now indicating the pipe index -- with respect to each flow rate, subject to the set of continuity equations (conservation laws) at each node.  This ``pipe network MinEP principle'' is examined in detail here. 

\item A separate MaxEPMod principle has recently been given for electrical circuit analysis \cite{Zupanovic_etal_2004, Botric_etal_2005, Christen_2006, Bruers_etal_2007a}. In this method, \eqref{eq:EP} is maximised -- again for pipe index $j$ -- with respect to the flow rates, subject to the constraint that the heat production equals the rate of work on the system at steady state. Some workers \cite{Bruers_2007c} have confused this approach with the {Ziegler-Paltridge} MaxEP principle, when in fact it is closer in character to the pipe network MinEP principle. This approach is examined herein.

\item A set of methods for pipe network analysis have been proposed based on maximisation of the Shannon entropy (MaxEnt), as defined in several ways \cite{Awumah_etal_1990, Tanyimboh_T_1993, deSchaetzen_etal_2000, Formiga_etal_2003, Ang_J_2003, Setiadi_etal_2005}. Although developed more for the purpose of reliability engineering and optimal network design (operational research), these methods are commented upon here.

\end{list}

%\noindent The present analysis will examine the above principles in single pipes and parallel pipe networks. The more difficult problem of flow in a generalised network -- in which not all nodes are connected, and with external flows to or from each node -- will be examined elsewhere. 

%############################################################################
\section{\label{pipeflow}Principles of Fluid Flow in Pipes} 
%% ############################################################################
%
We now consider one-dimensional, isothermal\footnote{Frictional flow causes heating, and so cannot strictly be isothermal. In keeping with most fluid mechanics references, the resulting temperature change in incompressible flow is here considered negligible \citep[see][p 37]{Bejan_1982}.}, constant composition, steady-state flow of an incompressible fluid between sections $A$ and $B$ in a single pipe, which satisfies the following energy conservation (augmented Bernoulli) equation \cite{Street_etal_1996, Schlichting_2001}:
\begin{equation}
z_A + \frac{p_A}{\rho g} + \alpha_A \frac{U_A^2}{2g} = z_B + \frac{p_B}{\rho g} + \alpha_B \frac{U_B^2}{2g} + H_L
\label{eq:Bernoulli}
\end{equation}
where $z_i$, $p_i$ and $U_i$ are respectively the elevation, pressure and mean velocity of the flow at section $i$, $\alpha_i$ is the kinetic energy correction factor at section $i$ (which corrects for non-one-dimensional effects), $\rho$ is the fluid density, $g$ is the acceleration due to gravity, and $H_L=p_L/\rho g$ is the head loss between sections $A$ and $B$, equal to the pressure loss $p_L$ per unit specific weight of the fluid. Note that $p_L$ does not simply give the pressure drop $p_B-p_A$, but represents the energy per volume lost due to friction between $A$ and $B$. We here consider only long, straight pipes of uniform circular section, of pipe diameter $d$, mean velocity $U_A=U_B=U$ and volumetric flow rate $Q=\pi d^2 U/4$. From the second law of thermodynamics, in all cases $H_L > 0$ in the direction of flow. However, to allow either choice of flow direction, it is convenient to consider $H_L >0$ (head loss) for $Q>0$ and $H_L<0$ (head gain) for $Q<0$. The head loss is given by the Darcy-Weisbach equation:
%\pagebreak
\begin{equation}
H_L = \frac{f L}{2gd} {U |U|} = \frac{8 f L}{\pi^2 g d^5} Q |Q| %= X Q^2
\label{eq:DW}
\end{equation}
where $f$ is the Darcy friction factor, a dimensionless drag force (= 4 times the Fanning friction factor), and $L$ is the pipe length. The absolute value signs in \eqref{eq:DW} allow for reverse flow.  

\begin{figure}[t]
%\begin{center}
\setlength{\unitlength}{0.6pt}
  \begin{picture}(500,275)
   \put(10,0){\includegraphics[width=80mm]{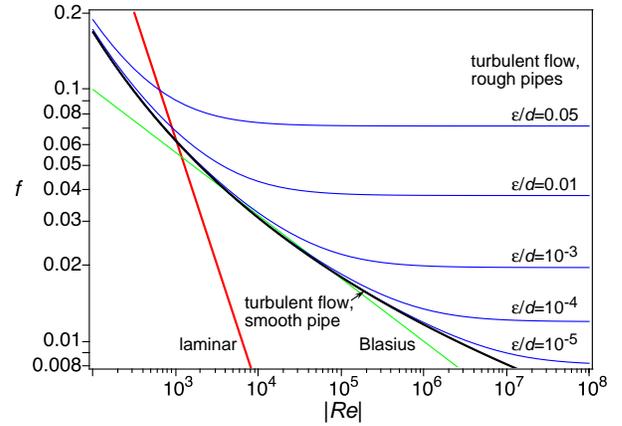}}
  \end{picture}
%\end{center}
\caption{Plots of $f(Re)$ for laminar flow \eqref{eq:f_lam} and turbulent flow in smooth \eqref{eq:f_sm} and rough \eqref{eq:f_turb} pipes (``Moody diagram''). The Blasius correlation \eqref{eq:f_Blas} is also shown.}
\label{fig:Moody}
\end{figure}

In laminar flow, the friction factor and head loss can be derived analytically as \cite{Street_etal_1996, Schlichting_2001}:
\begin{gather}
f_{lam} = \frac{64}{|Re|}
\label{eq:f_lam}
\\
H_{L,lam} = \frac{128 \mu L}{\pi \rho g d^4} Q
\label{eq:HL_lam}
\end{gather}
in which the Reynolds number of flow is given by:
\begin{equation}
Re = \frac{\rho d U}{\mu} = \frac{4 \rho Q}{\pi \mu d}
\label{eq:Re}
\end{equation}
(allowing $Re \gtrless 0$) and $\mu$ is the dynamic viscosity. Eq.\ \eqref{eq:HL_lam} is the well-known Hagen-Poiseuille equation. For turbulent flow, no theoretical solution is available, and so the friction factor is usually correlated using the following {semi-}empirical relations, respectively for flow in a smooth pipe, in a fully rough pipe or in general \cite{Street_etal_1996, Schlichting_2001}:
\begin{gather}
\frac{1}{\sqrt{f_{sm}}} = 2.0 \log_{10} (|Re| \sqrt{f_{sm}}) - 0.8
\label{eq:f_sm}
\\
\frac{1}{\sqrt{f_{full}}} = - 2.0 \log_{10}  \Bigl( \frac {\epsilon}{d} \Bigr) + 1.14
\label{eq:f_full}
\end{gather}
\vspace{-10pt}
\begin{multline}
\frac{1}{\sqrt{f_{turb}}} + 2.0 \log_{10}  \Bigl( \frac {\epsilon}{d} \Bigr) = 
\\
1.14 - 2.0 \log_{10} \Biggl [ 1 + \frac {9.28} {|Re| \Bigl( \dfrac {\epsilon}{d} \Bigr) \sqrt{f_{turb}}} \Biggr ]
\label{eq:f_turb}
\end{multline}
where $\epsilon$ is the equivalent sand roughness, a measure of the pipe wall roughness.  Eqs.\ \eqref{eq:f_sm}-\eqref{eq:f_turb} %can be converted to explicit functions of $f$ using the LambertW function \cite{***}, and the three equations 
can be solved in conjunction with \eqref{eq:DW} to give the turbulent head loss.  However, for engineering purposes, they are usually applied graphically using a plot of $f$ against $|Re|$ (the ``Moody diagram''). This diagram, constructed from eqs.\ \eqref{eq:f_lam} and \eqref{eq:f_sm}-\eqref{eq:f_turb}, is shown in Figure \ref{fig:Moody}. As shown, the flow remains laminar up to approximately $|Re| \approx 2100$, at which a transition takes place to a flow of fluctuating laminar and turbulent character. {Typically,} by $|Re| \approx 4000$ the flow has become entirely turbulent, {although laminar flow can be prolonged to $|Re| > 10000$ in careful, vibration-free experiments \cite{Bird_etal_2006, Munson_etal_2009}. Once the flow becomes turbulent, for non-zero $\epsilon$ it follows one of the curves $f(|Re|,\epsilon)$ \eqref{eq:f_turb} with increasing $|Re|$} until it reaches a condition of ``fully developed turbulent flow" \eqref{eq:f_full}, whereupon $f$ becomes independent of $|Re|$ and is a function only of $\epsilon/d$. 

Flow in a smooth pipe ($\epsilon/d \to 0$) follows the curve given by \eqref{eq:f_sm}. This is often correlated using the simpler Blasius relation, valid over $4000 \lesssim |Re| \lesssim 10^5$ \cite{Street_etal_1996, Munson_etal_2009}:
\begin{gather}
f_{Blas} = \frac{0.316}{|Re|^{0.25}}
\label{eq:f_Blas}
\end{gather}
which in \eqref{eq:DW} gives: 
\begin{gather}
H_{L,Blas} = \frac{1.787 \mu^{0.25} L}{\pi^{1.75} g d^{4.75} \rho^{0.25}}  Q |Q|^{0.75} 
\label{eq:HL_Blas}
\end{gather}
Eq.\ \eqref{eq:f_Blas} is also shown in Figure \ref{fig:Moody}.  In hydraulics, the laminar, Blasius and fully developed turbulent head loss curves are often expressed using the power law:
\begin{gather}
H_{L,\alf} = X Q |Q|^{\alf-1}   
\label{eq:HL_power}
\end{gather}
where the coefficient $X$ is a function only of the pipe and fluid properties, whilst $1 \le \alf \le 2$ expresses the transition between laminar ($\alf=1$) and fully developed turbulent flow ($\alf=2$).

%####################
\section{\label{single}Entropy Production in a Single Pipe} 
%% ############################################################################

\begin{figure}[!]
%\begin{center}
\setlength{\unitlength}{0.6pt}
  \begin{picture}(500,580)
   \put(0,310){(a)}
   \put(0,300){\includegraphics[width=85mm]{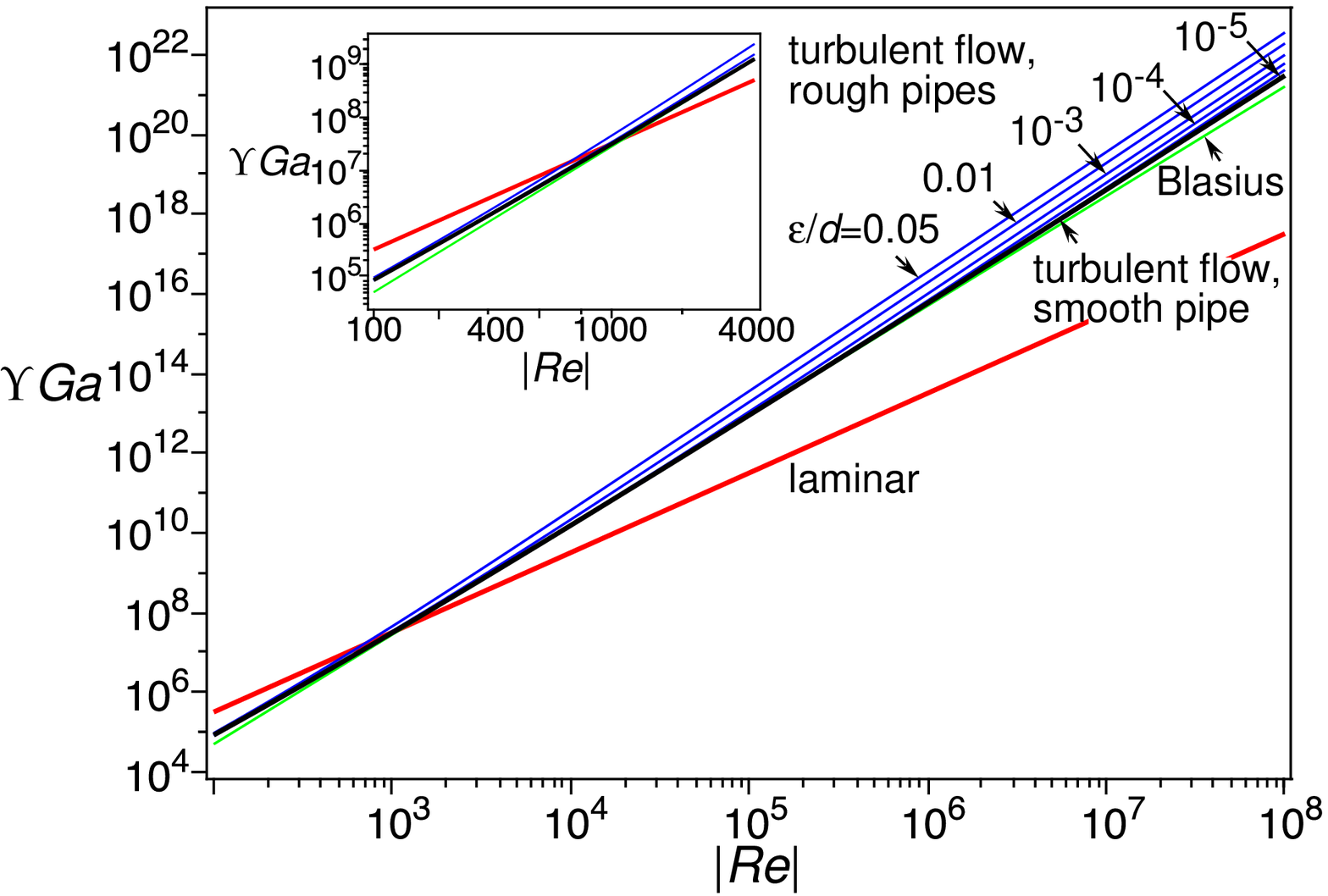}}
   \put(0,10){(b)}
   \put(0,0){\includegraphics[width=85mm]{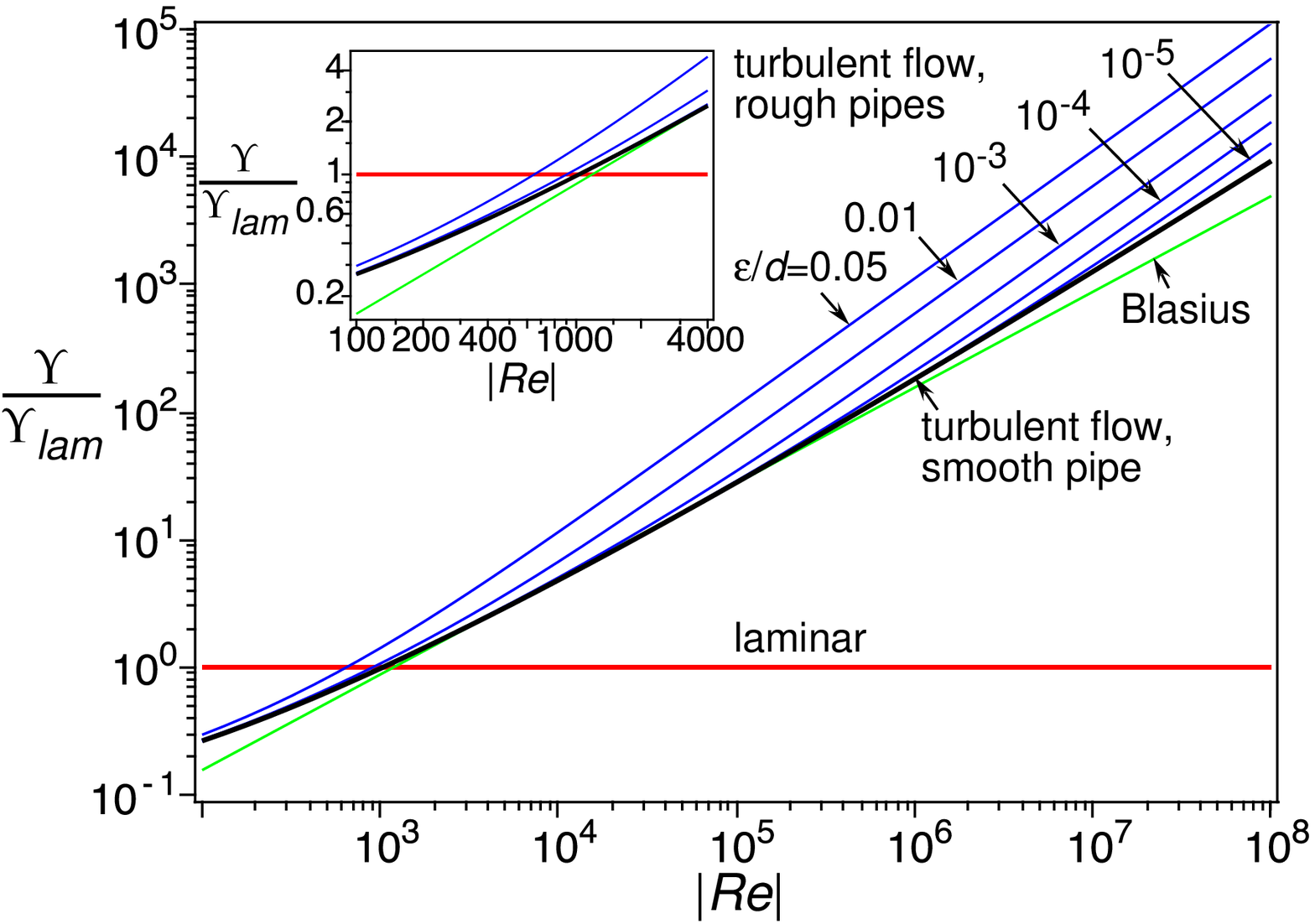}}
  \end{picture}
%\end{center}
\caption{Plots of (a) $\E Ga$ and (b) $\E/\E_{lam}$ against $Re$ for the same curves as in Figure \ref{fig:Moody}, calculated using \eqref{eq:EPdim}. The insets show the transition region. }
\label{fig:Moody_EP}
\end{figure}

In all the above cases, the rate of entropy production $\dot{\sigma}$ per unit specific weight of fluid for incompressible, steady-state flow in a single pipe, at constant temperature $T$ (equal to the power loss $P_L$ per unit specific weight), is given by \cite{Bejan_1982, Bejan_1996, Bejan_2006}:
\begin{equation}
\frac{P_L}{\rho g} = \frac{\dot{\sigma} T}{\rho g} = |H_L Q| \ge 0
\label{eq:EP_pipe}
\end{equation}
The dimensions of \eqref{eq:EP_pipe} suggest use of the following dimensionless group (normalised rate of entropy production or power loss per unit length of pipe):
\begin{equation}
\E = \frac{4 P_L}{\pi d \mu g L} = \frac{4 \dot{\sigma} T}{\pi d \mu g L} = \Bigl | \frac {H_L}{L} Re \Bigr | = \frac{1}{2} f \frac{|Re|^3}{Ga} \ge 0
\label{eq:EPdim}
\end{equation}
where the Galileo number $Ga = \rho^2 g d^3 / \mu^2$, which expresses the ratio of gravity and inertial to viscous forces, depends only on the pipe and fluid properties \cite{Pavlov_etal_1979, Niven_2002}. To the author's knowledge, the composite group $\E$ \eqref{eq:EPdim} has not been suggested previously. Several workers \cite{Schlichting_2001, PaulusJr_2000, PaulusJr_G_2004} use the group $f Re^2$ or its square root; however, this only correlates with the dimensionless head loss $H_{L}/L = \frac{1}{2} f Re |Re|/Ga$, not the entropy production.

Plots of the normalised entropy production $\E Ga$ against $|Re|$ for the various curves for laminar and turbulent flow are shown in Figure \ref{fig:Moody_EP}(a). Plots of the ratio $\Upsilon/\Upsilon_{lam}$ against $|Re|$ are given in Figure \ref{fig:Moody_EP}(b). As evident, for $|Re| < 1034$ the entropy production by laminar flow in a smooth pipe exceeds that of turbulent flow, but the latter increases more rapidly with $|Re|$; for $|Re|=10^4$, $10^6$ and $10^8$, the respective ratios are $\E/\E_{lam}=4.82$, $182.0$ and $9282$. The known onset of the transition region at $|Re| \approx 2100$ is {of the order of (a little higher than)} the crossover point {(see later discussion)}. Turbulent flow in rough pipes follows similar trends, with even higher entropy production than a smooth pipe. Yet, for all Reynolds numbers, the laminar solution \eqref{eq:f_lam}-\eqref{eq:HL_lam} is a valid steady-state solution, being the analytical solution of the Navier-Stokes equations. The laminar and turbulent curves in Figures \ref{fig:Moody_EP}(a)-(b) therefore represent two alternative steady-state solutions, from which the system appears {-- for the most part --} to select the one with the highest entropy production. Thus although not constituting a proof, Figures \ref{fig:Moody_EP}(a)-(b) provide strong evidence for the action of the {Ziegler-Paltridge} MaxEP principle (\S\ref{EP_principles}) in providing a driving force for the selection of laminar or turbulent flow. 

The above argument was given by Paulus Jr \cite{PaulusJr_2000, PaulusJr_G_2004} and Martyushev \cite{Martyushev_2007}, but presented graphically in terms of the friction factor instead of the entropy production.  Note that the trends of the entropy production curves in Figure \ref{fig:Moody_EP}(a) are remarkably similar to those of other ``complex systems'' which undergo a critical transition between linear and non-linear dissipation mechanisms. These include (i) heat transfer through a convective fluid, for which $\dot{\sigma}$ dramatically increases with Rayleigh number at the transition from heat diffusion to convection \cite{Schneider_K_1994}; and (ii) chemical degradation in an ecosystem, for which $\dot{\sigma}$ dramatically increases with chemical affinity at the transition from chemical to biological metabolic processes \cite{Bruers_M_2007}.  Clearly, the {Ziegler-Paltridge} MaxEP concept unites many quite different disciplines (see \cite{Niven_MEP}).

{Examining the discrepancy between the typical onset of turbulent flow at $|Re| \approx 2100$ and the crossover point of $|Re|=1034$ in Figures \ref{fig:Moody_EP}(a)-(b) (with lower values in rough pipes), two explanations are forthcoming. Firstly, the discrepancy may result from the use of empirical relations for the turbulent friction factor \eqref{eq:f_sm}-\eqref{eq:f_turb}, which are not fitted to nor intended to be applied to the transition region. Secondly, the discrepancy may be due to the formation of ``metastable'' laminar flow at higher Reynolds numbers than would otherwise be expected from the entropy production. This phenomenon is well known, as demonstrated by the sudden transition to turbulence during flow in a pipe, in contrast to the more gradual transition seen in flow in packed beds and many external and boundary layer flows \cite{Bird_etal_2006, Munson_etal_2009, Street_etal_1996, Schlichting_2001, Niven_2002}. Also, as mentioned earlier, laminar flow can persist at higher $|Re|$ in careful experiments, even above $10000$, but is unstable to vibrational disturbance \cite{Bird_etal_2006, Munson_etal_2009}. This situation is somewhat analogous to the existence of a metastable supersaturated solution of a salt, despite a free energy gradient which should drive its precipitation; in such cases, perturbation of the system (such as tapping the container wall) can suffice to drive it to equilibrium. No-one discards thermodynamics because of this exceptional phenomenon. The relevant question here is not whether metastable laminar flow can occur at high $|Re|$, but what is the minimum critical $|Re|$ for the onset of turbulence?  Some research in this direction has been undertaken by Benhamou et al. \cite{Benhamou_etal_2004} \citep[see also][]{Martyushev_2007}, who reported significantly lower critical $|Re|$ of around 1200-1500 in a horizontally oscillating pipe, and by Li et al. \cite{Li_etal_2009}, who found the critical $|Re|$ can drop to 1600-2000 due to pipe entrance effects.}

{For completeness}, we also consider the (somewhat absurd) variation of the pipe entropy production with respect to the flow rate $Q$.  Taking \eqref{eq:EP_pipe} as the objective function $F$, using the power-law relation \eqref{eq:HL_power} for the head loss, differentiation gives:
\begin{align}
%\begin{split}
\frac {\partial F}{\partial Q} &= (\alf+1) X Q |Q|^{\alf-1} = (\alf+1) H_{L}
%\end{split}
%\raisetag{20pt}
\label{eq:diff_EP_single}
\end{align}
This is non-zero unless $Q=0$, hence the steady state is not an extremum in the entropy production with respect to $Q$.  This does not affect the above-mentioned MaxEP principle, which applies with respect to the selection of flow regime and is unrelated to the flow rate.  However, \eqref{eq:diff_EP_single} indicates that the pipe network MinEP principle -- which involves the variation with respect to flow rates -- does not apply to fluid flow in a single pipe.  The validity of the MaxEPMod principle of \Zup and co-workers for flow in a single pipe is deferred to \S\ref{parallel_MaxEPMod}.

%############################################################################
\section{\label{parallel}Entropy Production in Parallel Pipes} 
%\subsection{\label{closed}Closed-Form Solution} 
\subsection{\label{fixedQ}Fixed Flow Rate Systems and the Pipe Network MinEP Principle}
%% ############################################################################
%
\begin{figure}[t]
%\begin{center}
\setlength{\unitlength}{0.6pt}
  \begin{picture}(500,130)
   \put(75,0){\includegraphics[width=55mm]{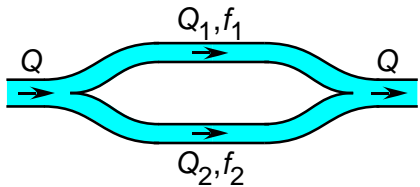}}
  \end{picture}
%\end{center}
\caption{A simple parallel pipe network with fixed total flow rate $A$.}
\label{fig:parallel}
\end{figure}

We now turn attention to the simple pipe geometry of Figure \ref{fig:parallel}, in which the flow $Q$ is forced to subdivide between two parallel pipes, of flow rates $Q_1$ and $Q_2$, which then rejoin. The pipes have different flow characteristics.  We here neglect head losses associated with the pipe junctions and fittings. Given the pipe and fluid properties, how will the flow divide between $Q_1$ and $Q_2$?  In fact, a closed-form solution exists. Firstly, the flows are subject to continuity (conservation of fluid) equations at each node, in this case identical at both nodes \cite{Street_etal_1996}:
\begin{equation}
Q-Q_1-Q_2=0
\label{eq:cty_case0}
\end{equation}
This is equivalent to Kirchhoff's current law in electrical circuit analysis, taking as analogues the current $I_j$ and flow rate $Q_j$ \cite{Stephen_1960, Morris_1993}. Secondly, the system is considered to satisfy the condition that at steady state, there should be no circular flow around any loop, or alternatively, that the head losses along all paths between two nodes must be the same  \cite{Street_etal_1996}:
\begin{equation}
H_{L1}-H_{L2}=0
\label{eq:loop_case0}
\end{equation}
This is a non-linear variant of Kirchhoff's voltage law in electrical circuit analysis \cite{Stephen_1960, Morris_1993}, taking the voltage drop $V_j$ and head loss $H_{Lj}$ as analogues, consistent with the notion that each node should be at a constant steady-state piezometric head (potential). Substituting \eqref{eq:DW} for the head losses in \eqref{eq:loop_case0}, using the friction factor \eqref{eq:f_lam}, \eqref{eq:f_sm}, \eqref{eq:f_full} or \eqref{eq:f_turb} for the applicable flow regime, yields two equations in the two unknowns $Q_1$ and $Q_2$. For example, using the power law relation \eqref{eq:HL_power}, \eqref{eq:loop_case0} reduces to 
\begin{equation}
X_1 Q_1 |Q_1|^{\alf-1}-X_2 Q_2 |Q_2|^{\alf-1}=0
\label{eq:loop2_case0}
\end{equation}
Knowing $\alf$ and $X_j$ for $j \in \{1,2\}$, \eqref{eq:cty_case0} and \eqref{eq:loop2_case0} can be solved numerically for $Q_1$ and $Q_2$. If the flow regime is not {\it a priori} known, the solution requires an iterative selection scheme based on calculation of $Re_j$ and $X_j$. 

%%%
%\subsection{\label{parallel_MinEP}Pipe Network MinEP Principle} 
%%%

Although a closed-form solution exists, its solution becomes inconvenient for larger pipe networks, requiring continuity equations for all nodes and head balance equations for all loops.  Usually, the problem is over-constrained (more equations than unknowns), causing difficulties with numerical solution.  However, there exists a quite different approach, which appears to have been almost completely overlooked in the engineering literature, except for two studies by Paulus Jr \cite{PaulusJr_2000, PaulusJr_G_2004}. A linear variant of the method, for electrical circuit analysis, is slightly better known \cite{Jeans_1966, Landauer_1975, Jaynes_1980, Kondepudi_P_1998}. The method involves constrained minimisation of the total entropy production at constant $T$ (equal to the total power loss) within the pipe system:
\begin{equation}
\frac{P_L}{\rho g}  = \frac{\dot{\sigma} T}{\rho g} = \sum\limits_{j=1}^J |H_{Lj} Q_j|
\label{eq:EP_network}
\end{equation}
with respect to the flow rates, subject to the node continuity constraints:
\begin{equation}
Q_{n} + \sum\limits_{j=1}^J Q_{jn} =0, \qquad n=1,...,N
\label{eq:cty_gen}
\end{equation}
where $j$ is the pipe index, $J$ is the number of pipes, $n$ is the node index, $N$ is the number of nodes, $Q_{jn}$ is the flow rate from pipe $j$ into node $n$ and $Q_{n}$ is the external inflow into node $n$. %This MinEP method is quite different to the other MinEP methods discussed in \S\ref{EP_principles}, such as of Prigogine.

It is worth examining the validity of this method for parallel pipe networks.  For the simple example in Figure \ref{fig:parallel}, assuming power law behaviour \eqref{eq:HL_power} with constant $\alf$, proof is relatively straightforward.  From \eqref{eq:EP_network}:
\begin{equation}
\frac{P_L}{\rho g} = \frac{\dot{\sigma} T}{\rho g} = |H_{L1} Q_1| + |H_{L2} Q_2|
\label{eq:EP_case0}
\end{equation}
Substituting for the head losses \eqref{eq:HL_power} and $Q_2$ \eqref{eq:cty_case0} gives the constrained objective function:
\begin{equation}
F %= \frac{P_L}{\rho g} \biggr|_{Q=Q_1+Q_2} 
= X_1 |Q_1|^{\alf+1} + X_2 |Q-Q_1|^{\alf+1}
\label{eq:EP_case0b}
\end{equation}
Differentiation yields:
\begin{align}
\begin{split}
\frac {\partial F}{\partial Q_1} &= (\alf+1) [X_1 Q_1 |Q_1|^{\alf-1} - X_2 (Q-Q_1) |Q-Q_1|^{\alf-1}]
\\
&= (\alf+1) [X_1 Q_1 |Q_1|^{\alf-1} - X_2 Q_2 |Q_2|^{\alf-1}]
\\
&= (\alf+1) [H_{L1}-H_{L2}]
\end{split}
\raisetag{24pt}
\label{eq:diff_EP_case0}
\end{align}
where \eqref{eq:HL_power} and \eqref{eq:cty_case0} are again used. For steady state flow, the last form of \eqref{eq:diff_EP_case0} must be zero, hence the steady state is an extremum. Differentiating a second time gives:
\begin{align}
\begin{split}
\frac {\partial^2 F}{\partial Q_1^2} &=\alf (\alf+1) [X_1 |Q_1|^{\alf-1} + X_2 |Q-Q_1|^{\alf-1}]
\\
&=\alf (\alf+1) [X_1 |Q_1|^{\alf-1} + X_2 |Q_2|^{\alf-1}]
\\
&=\alf (\alf+1) \Bigl[ \frac{H_{L1}}{Q_1} + \frac{H_{L2}}{Q_2} \Bigr]
\end{split}
\label{eq:diff2_EP_case0}
\end{align}
again using \eqref{eq:HL_power} and \eqref{eq:cty_case0}. Since the terms $H_{Lj}/Q_j$ are {\it always} positive at steady state irrespective of flow direction, ${\partial^2 F}/{\partial Q_1^2}>0$ for $\alf \ge 1$, and so the extremum is indeed a minimum.  A related proof, only for the laminar case, was given by Paulus Jr \cite{PaulusJr_2000}. Since this method uses the node continuity constraint \eqref{eq:cty_case0}, and from \eqref{eq:diff_EP_case0}, recovers the loop head constraint \eqref{eq:loop_case0}, it is identical to the closed-form solution. 

The above proof can be extended to any number $J \ge 2$ of parallel pipes with common $\alf$ between two nodes, in which case one obtains $J-1$ derivatives with respect to $Q_{k}$, each of which is a local minimum. The steady state position is therefore a global minimum.  The MinEP and closed-form solutions are again equivalent. %For compressible flows, it is possible to recast the theorem in terms of the mass flow rates $\dot{m}_j$ rather than $Q_j$; however, an additional equation of state (between pressure and density) is necessary for solution closure.

Difficulties arise, however, with the pipe network MinEP principle for other forms of the friction factor $f(Re)$. Consider the case of two parallel pipes (Figure \ref{fig:parallel}) with constant but different power-law coefficients $\alf_1$ and $\alf_2$ in \eqref{eq:HL_power}. From \eqref{eq:cty_case0}
 and \eqref{eq:EP_network}:
\begin{equation}
F = X_1 |Q_1|^{\alf_1+1} + X_2 |Q-Q_1|^{\alf_2+1}
\label{eq:EP_case0_alphas}
\end{equation}
from which:
\begin{align}
\begin{split}
\frac {\partial F}{\partial Q_1} &= (\alf_1+1) X_1 Q_1 |Q_1|^{\alf_1-1}  
\\
& \quad - (\alf_2+1) X_2 (Q-Q_1) |Q-Q_1|^{\alf_2-1}
\\
&= (\alf_1+1) X_1 Q_1 |Q_1|^{\alf_1-1} - (\alf_2+1) X_2 Q_2 |Q_2|^{\alf_2-1}
\\
&= (\alf_1+1) H_{L1} - (\alf_2+1) H_{L2}
\end{split}
\raisetag{50pt}
\label{eq:diff_EP_case0_alphas}
\end{align}
This is certainly not zero unless (trivially) $a_1=a_2$ or the pipes carry flows with $H_{L1} \ne H_{L2}$, an unstable situation. In consequence, the pipe network MinEP principle does not apply to parallel pipes of different $\alf_j$.  Indeed, as proven in Appendix \ref{Apx_nonpower}, it also fails for parallel pipe networks with non-power-law $f(Re)$ relations such as those in \eqref{eq:f_sm} or \eqref{eq:f_turb}. Thus although it is quite appealing to consider the pipe network MinEP principle as a fundamental principle of fluid mechanics, as advocated by Paulus Jr \cite{PaulusJr_2000, PaulusJr_G_2004}, it is not.  A similar conclusion was reached by Landauer \cite{Landauer_1975} with regards to electrical networks, for which the MinEP principle applies to circuits containing resistors, but not those with inductances. A further difficulty arises for variable temperature flows, which do not satisfy any MinEP principle, a criticism also made by several authors \cite{Landauer_1975, Jaynes_1980, Martyushev_etal_2007} in reference to electrical networks. The term ``minimum entropy production principle'' is therefore rather misleading, since the principle uncovered here is more accurately described as a ``minimum power loss'' principle, applicable only to the restricted set of isothermal flows in constant $\alf$ pipe networks.

It is important to clarify the differences between the pipe network MinEP principle and that of Prigogine \cite{Prigogine_1967, Prigogine_1980, Prigogine_S_1984, Kondepudi_P_1998}. Firstly, as noted earlier, the former involves the constrained minimisation of a sum of entropy production terms for $J \ge 2$ pipes, within which the flow rates $Q_j$ are not independent, whereas Prigogine's theorem involves a sum of entropy production terms for $J \ge 2$ independent physical phenomena. %(Indeed, as noted earlier \cite{Zupanovic_etal_2004, Martyushev_etal_2007} but is still not appreciated by some authors \cite{Kondepudi_1998, Bruers_2007c}, Prigogine's theorem cannot be applied to a single transport phenomenon such as isothermal fluid flow.)  
Secondly, the present theorem applies to both linear ($\alf=1$) and non-linear ($1 < \alf \le 2$) functional relationships $H_{Lj}(Q_j)$, albeit only for the restricted case of systems with common $\alf$. This is in contrast with Prigogine's method, which requires linear (Onsager) relations between the forces and fluxes.  However, the present theorem is in some sense allied to Prigogine's, in that both emerge -- in some circumstances -- as a simple (almost trivial \cite{Jaynes_1980}) consequence of stationarity of the steady state compared to non-steady-state choices of the flow rates. As such, the pipe network MinEP principle -- when it applies - relates only to the dynamic stability of the steady state, and does not have any bearing on the selection of the steady state if there exist multiple steady state solutions.

%%%
\subsection{\label{parallel_MaxEPMod}Fixed Head Systems and the MaxEPMod Principle} 
%%%

\begin{figure}[t]
%\begin{center}
\setlength{\unitlength}{0.6pt}
  \begin{picture}(500,360)
   \put(0,185){\bf{(a)}}
   \put(0,185){\includegraphics[width=85mm]{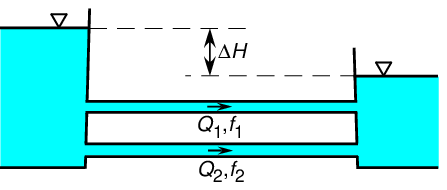}}
   \put(0,10){\bf{(b)}}
   \put(150,0){\includegraphics[width=25mm]{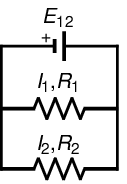}}
  \end{picture}
%\end{center}
\caption{Network diagrams for the MaxEPMod method: (a) parallel pipes between reservoirs, and (b) analogous electrical circuit.}
\label{fig:Zup}
\end{figure}

We now consider a new method for electrical circuit analysis given by \Zup and co-workers \cite{Zupanovic_etal_2004, Botric_etal_2005, Christen_2006, Bruers_etal_2007a}, which is here extended to a parallel pipe network. A slightly different flow configuration is required, as represented in Figure \ref{fig:Zup}(a), in which the system is constrained by a fixed change in head $\Delta H$ (assuming infinite reservoirs) rather than a fixed total flow rate $Q$. The analogous electrical network is shown in Figure \ref{fig:Zup}(b).  From the head loop principle, the system yields two loop equations:
\begin{gather}
\Delta H = H_{Lj}, \qquad j=1,2
\label{eq:loop_rels}
\end{gather}
which can be solved for $Q_1$ and $Q_2$. The MaxEPMod method involves maximisation of the entropy production multiplied by constant $T$ (equivalent to the heat production ${\dot{q}}$) \eqref{eq:EP_case0}, subject to the constraint that the heat production is equal to the work expenditure $\dot{w}$: 
\begin{equation}
{\dot{w}}  = {\rho g} |\Delta H \, Q| = {\rho g} |\Delta H| \, |Q_1 + Q_2| 
\label{eq:work_network}
\end{equation}
Applying the calculus of variations, assuming power-law pipes \eqref{eq:HL_power} of constant $\alf$, gives the Lagrangian:
\begin{align}
\begin{split}
G &= \frac{\dot{q}}{\rho g} + \lambda \bigl(\frac{\dot{w}}{\rho g} - \frac{\dot{q}}{\rho g} \bigr) = (1-\lambda) \frac{\dot{q}}{\rho g} + \lambda \frac{\dot{w}}{\rho g} 
\\
&= (1-\lambda) \bigl( |H_{L1} Q_1| + |H_{L2} Q_2| \bigr) + \lambda |\Delta H| \, |Q_1 + Q_2| 
\\
&= (1-\lambda) \bigl( X_1 |Q_1|^{\alf+1} + X_2 |Q_2|^{\alf +1} \bigr) + \lambda |\Delta H| \, |Q_1 + Q_2| 
\end{split}
\label{eq:Zup_Lagr}
\raisetag{50pt}
\end{align}
where $\lambda$ is the Lagrangian multiplier. Setting the variation with respect to each $Q_j$ to zero gives:
\begin{multline}
\frac{\partial G}{\partial Q_j} = (1-\lambda)(\alf+1) X_j Q_j |Q_j|^{\alf-1} 
\\ + \lambda |\Delta H| \sign(Q_1+Q_2) = 0
\label{eq:Zup_diffL}
\end{multline}
which, from \eqref{eq:HL_power}, gives:
\begin{equation}
\lambda = \frac{(\alf+1) H_{Lj}}{(\alf+1) H_{Lj} - |\Delta H| \sign (Q)}, \quad \forall j
\label{eq:Zup_lambda}
\end{equation}
For $Q \gtrless 0$, the extremum always coincides with the loop equations $\Delta H = |\Delta H| \sign (Q)=H_{Lj}, \forall j$, if we set $\lambda = (\alf+1)/\alf$. This reduces to the value of $\lambda=2$ given previously for electrical networks ($\alf=1$) \cite{Zupanovic_etal_2004, Botric_etal_2005}. Since the method recovers the two loop constraints -- the node continuity now being irrelevant -- it is identical to the closed-form solution. Taking the second derivative gives, for $Q_j \ne 0$:
\begin{align}
\begin{split}
\frac{\partial^2 G}{\partial Q_j^2} &=(1-\lambda) \alf (\alf+1) X_j |Q_j|^{\alf-1}
\\
&= - (\alf+1) X_j |Q_j|^{\alf-1} = - \frac{(\alf+1) H_{Lj}}{Q_j}
\end{split}
\label{eq:Zup_diff2L}
\end{align}
Again $H_{Lj}/Q_j>0, \forall j$ and so ${\partial^2 G}/{\partial Q_1^2}<0$ for $\alf \ge 1$, hence the extremum is a maximum. By symmetry, the above analysis can be generalised to any number of parallel pipes, and indeed, even to a single pipe.

Difficulties again emerge, however, in other situations. For pipes with different $\alf_j$, one again obtains \eqref{eq:Zup_diffL}-\eqref{eq:Zup_lambda} but expressed in terms of $\alf_j$ instead of $\alf$. No common value of $\lambda$ exists which can satisfy all loop constraints. More broadly, as proven in Appendix \ref{Apx_nonpower2}, it also fails for parallel pipe networks with non-power-law $f(Re)$ relations. Furthermore, just like the pipe network MinEP principle, the MaxEPMod approach fails for variable temperature systems, and so is more accurately described as a heat-work extremum principle\footnote{Note from the first line of \eqref{eq:Zup_Lagr}, knowing $\lambda>1$, the MaxEPMod principle could equally have been posed as a MinEP principle, subject to a constant work constraint, or a maximum work principle, subject to a constant heat production constraint.}.  The MinEP and MaxEPMod principles therefore have many features in common, and so can be regarded as kindred principles differing primarily in their choice of flow conditions.

%%%
\subsection{\label{parallel_MaxEnt}MaxEnt Methods} 
%%%

As mentioned earlier, several methods for pipe network analysis have been proposed based on maximisation of the Shannon entropy (MaxEnt) \cite{Awumah_etal_1990, Tanyimboh_T_1993, deSchaetzen_etal_2000, Formiga_etal_2003, Ang_J_2003, Setiadi_etal_2005}. Of these, most have been developed for reliability or sensitivity analysis, e.g.\ the method of Awumah and co-authors \cite{Awumah_etal_1990}, in which the pipe network of highest entropy is considered the most reliable.  Tanyimboh and Templeman \cite{Tanyimboh_T_1993} propose three separate MaxEnt methods for the analysis of separate types of pipe networks, with various definitions of probabilities and entropies.  Unfortunately, all three methods permit only one connection between each pair of nodes, and therefore cannot be applied to the parallel pipe problems considered here.  Furthermore, none of the methods incorporate frictional resistances for each pipe (such as $f_j$ or $X_j$) and so their results assume equivalence of all pipes, a very unrealistic situation. 

%%%
\subsection{\label{parallel_MaxEP}MaxEP Principle} 
%%%

For parallel pipe flow, we must not forget the action of the quite distinct {Ziegler-Paltridge} MaxEP principle (\S\ref{EP_principles}). %, which applies equally to single or parallel pipe flows.  Indeed, if flow in one pipe in a parallel network becomes turbulent, causing a higher head loss, the requirement of head loss balance \eqref{eq:loop2_case0} will tend to drive the other pipe flows to be turbulent. 
The interplay between this extremum principle and the MinEP or MaxEPMod principles is examined in the next section, by the analysis of examples for constant flow rate or constant head conditions.

%############################################################################
\section{\label{eg}Examples}
%##############################################
%\subsection{\label{eg1}Example 1}
%%#############################################

We now illustrate the various roles of the {Ziegler-Paltridge} MaxEP, pipe network MinEP and \Zup MaxEPMod principles in parallel pipe networks by two examples.  Example 1 is of two parallel pipes with fixed total flow rate $Q$ (Figure \ref{fig:parallel}), in which the pipes can experience either laminar flow \eqref{eq:f_lam}-\eqref{eq:HL_lam} or turbulent flow represented by the Blasius relation \eqref{eq:f_Blas}-\eqref{eq:HL_Blas}.  Since these relations conform to power-law behaviour \eqref{eq:HL_power}, the pipe network MinEP principle applies; we also examine the {Ziegler-Paltridge} MaxEP principle. The system is assumed to carry water, with properties $\rho=1000$ kg m$^{-3}$, $\mu = 10^{-3}$ Pa s, $g = 9.81$ m s$^{-2}$ and with both pipes of the same diameter $d_1=d_2 = 0.05$ m; the difference between pipe resistances is due to their different lengths, $L_1=50$ m and $L_2=150$ m. In consequence, the Reynolds number of each pipe $Re_j$ is proportional only to its flow rate $Q_j$ and independent of diameter $d_j$ \eqref{eq:Re}, whence $Q$ can be expressed in dimensionless form in terms of the composite Reynolds number $Re_{tot} = 4 \rho Q/ \pi \mu d = Re_1 + Re_2$. In each case, the laminar and turbulent steady-state solutions over a range of $Re_{tot}$ were separately computed by minimisation of the normalised entropy production \eqref{eq:EP_case0}, subject to the continuity constraint \eqref{eq:cty_case0}. Computations were conducted at 15-digit precision using the Minimize command in Maple 12 \cite{Waterloo_Maple}, which invokes a linear or nonlinear optimisation routine as needed; in all cases, convergence was attained within 4 iterations for both laminar and turbulent flow. Cross-checking of each solution against its closed-form counterpart \eqref{eq:cty_case0}-\eqref{eq:loop_case0} indicated flow rate errors of $< 10^{-19}$ m$^3$ s$^{-1}$ for laminar flow and $< 10^{-9}$ m$^3$ s$^{-1}$ for the Blasius relation.  

\begin{figure}[t]
%\begin{center}
\setlength{\unitlength}{0.6pt}
  \begin{picture}(550,600)
   \put(0,320){\includegraphics[width=85mm]{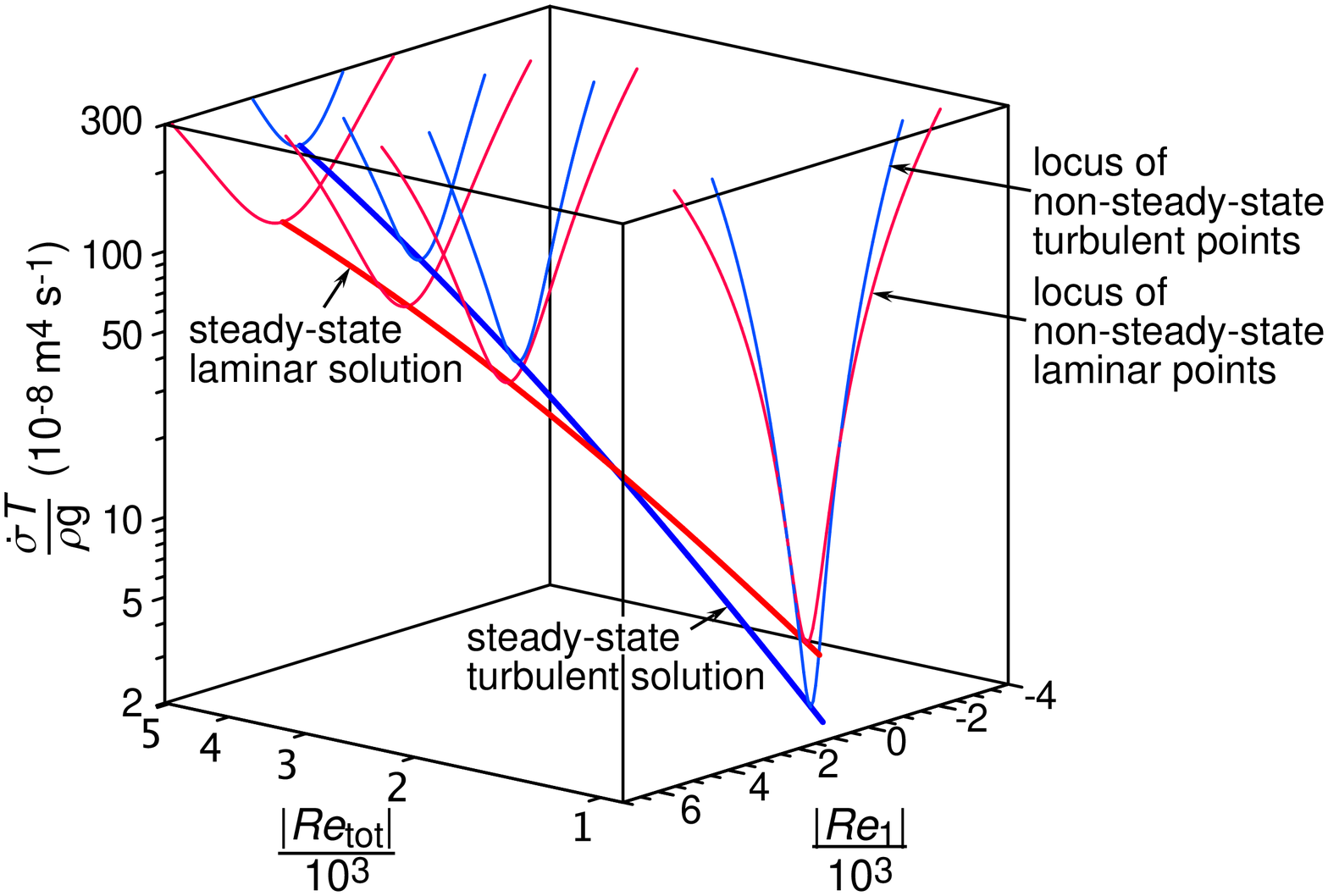}}
   \put(0,320){\bf{(a)}}
   \put(0,10){\includegraphics[width=85mm]{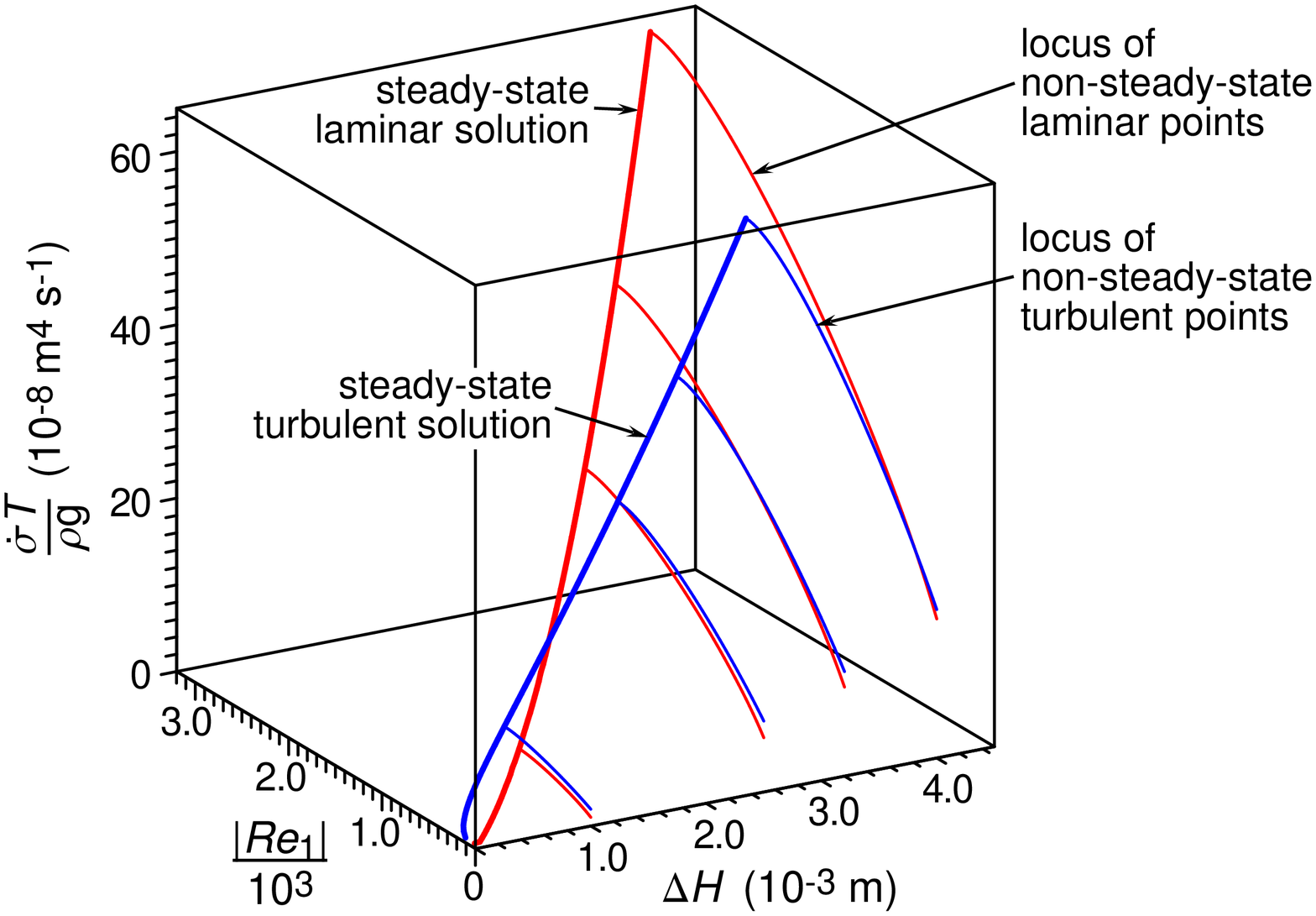}}
   \put(0,10){\bf{(b)}}
     \end{picture}
%\end{center}
\caption{Plots of normalised entropy production in parallel pipes for (a) constant $|Re_{tot}|$ (Example 1), showing action of {Ziegler-Paltridge} MaxEP and pipe network MinEP principles, and (b) constant $\Delta H$ (Example 2), showing action of inverted {Ziegler-Paltridge} MaxEP and \Zup MaxEPMod principles. For details of calculations, see \S\ref{eg}.}
\label{fig:3D}
\end{figure}

The calculated laminar and turbulent steady-state solutions for Example 1 are illustrated as solid lines in the three-dimensional plot of normalised entropy production against $Re_{tot}$ and $Re_1$ in Figure \ref{fig:3D}(a). As expected, the laminar curve is of higher entropy production at lower Reynolds numbers, but is overtaken by the turbulent curve at higher Reynolds numbers\footnote{Since the lines do not intersect, for equality of $\dot{\sigma}$ and satisfaction of the continuity and head loss constraints \eqref{eq:cty_case0}-\eqref{eq:loop_case0}, the system will exhibit a transition region over the range $Re_{tot} = 1585$ to $3417$, with turbulent flow in pipe 1 and laminar flow in pipe 2. In each case the transition to turbulence is predicted to occur in each pipe at $Re_j = 1189$. If the smooth curve \eqref{eq:f_sm} is used instead of the Blasius curve, the predicted transition values shift to $Re_{tot} = 1379$ to $3039$ and $Re_j = 1034$.}, consistent with the {Ziegler-Paltridge} MaxEP principle.  Superimposed on this plot are several subsidiary sets of curves for specific values of $Re_{tot}$, which illustrate the locus of non-steady-state solutions for laminar or turbulent flow, as obtained by simultaneous solution of \eqref{eq:EP_case0} and the continuity constraint \eqref{eq:cty_case0} for given values of $Re_{tot}$ and $Re_1$.  As evident, in every case - both for laminar and turbulent flow - the steady-state solution lies at the minimum of this locus. Figure \ref{fig:3D}(a) therefore illustrates the {\it simultaneous} action, in parallel pipe systems of fixed $Re_{tot}$ (or $Q$), of (i) the pipe network MinEP principle, which discriminates the steady-state position from the set of non-steady-state solutions for each flow regime and specified $Re_{tot}$, and (ii) the {Ziegler-Paltridge} MaxEP principle, which provides a means to determine the flow regime at a specified $Re_{tot}$. 

%%%%%%%%%%%%%%
%\subsection{\label{eg2}Example 2}
%%%%%%%%%%%%%

The next Example 2 also involves two parallel pipes, but with a fixed total head $\Delta H$ (Figure \ref{fig:Zup}(a)) rather than fixed $Q$. The fluid and pipe properties are otherwise identical to those in Example 1. We again consider laminar \eqref{eq:f_lam}-\eqref{eq:HL_lam} or Blasius \eqref{eq:f_Blas}-\eqref{eq:HL_Blas} flow, both of power law form, hence the MaxEPMod principle applies. In this case, the laminar and turbulent steady-state solutions over a range of $\Delta H$ were computed by maximisation of the normalised entropy production (or heat generation) \eqref{eq:EP_case0}, subject to the heat-work constraint, which from \eqref{eq:EP_case0} and \eqref{eq:work_network} can be expressed in the form:
\begin{equation}
%\frac {C(\alf) \; \mu^{2-\alf}}{\rho^{2-\alf} g d^{3+\alf}} (L_1 Q_1^{\alf+1}+L_2 Q_2^{\alf+1}) = |\Delta H| |Q_1 + Q_2|
X_1^{(\alf)} |Q_1|^{\alf+1}+X_2^{(\alf)} |Q_2|^{\alf+1} = |\Delta H| |Q_1 + Q_2|
\label{eq:heat_work_gen}
\end{equation}
where $\alf=1$ and $1.75$ respectively for the laminar and Blasius relations, and $X_j^{(\alf)}$ are constants for each specified pipe and flow regime. The Maximize routine in Maple 12 with 15 digit precision was used for computations; in this system, convergence required 30 to 50 iterations, with an error of $< 10^{-7}$ m$^3$ s$^{-1}$.  The calculated laminar and turbulent steady-state solutions are illustrated as solid lines in Figure \ref{fig:3D}(b), now as a function of $\Delta H$ and $Re_1$. In contrast to Example 1, the laminar curve is of lower entropy production than the Blasius curve at lower $\Delta H$, but overtakes it\footnote{Again the lines do not cross, forcing formation of a transition region over $\Delta H=0.00155$ to $0.00466$ m using the Blasius relation, or $\Delta H=0.00135$ to $0.00405$ m using the smooth relation. The total and pipe Reynolds numbers at the transition endpoints are identical to those for Example 1.} at higher $\Delta H$. The analysis therefore suggests the action of a MinEP principle for the selection of flow regime, quite opposite to the {Ziegler-Paltridge} MaxEP principle. However, under constant $\Delta H$ conditions, the flow rates are not comparable; for example, $\Delta H = 0.003$ m produces a total laminar flow of $Q = 1.20 \times 10^{-4}$ m$^3$ s$^{-1}$ ($Re_{tot} = 3066$) or a turbulent flow of only $1.04 \times 10^{-4}$ m$^3$ s$^{-1}$ ($Re_{tot} = 2659$). The analysis therefore confirms the argument of Paulus Jr and co-authors \cite{PaulusJr_2000, PaulusJr_G_2004} that under constant head (or pressure) conditions, the flow regime can be selected by a MinEP principle.  Since the transition values are identical to those given for Example 1 by the {Ziegler-Paltridge} MaxEP principle, it is a simple inversion of the latter, not an entirely new principle.

As in the previous example, non-steady-state curves for specific values of $\Delta H$ are also superimposed on the plot, calculated by simultaneous solution of \eqref{eq:EP_case0} and the heat-work constraint \eqref{eq:heat_work_gen} for given values of $\Delta H$ and $Q_1$. As evident, the steady-state solution lies at the maximum of this locus, both in laminar and turbulent flow. A curious feature of this system is that the subsidiary curves are one-sided, i.e., the non-steady-state flow rate $Q_1$ (or $Re_1$) cannot exceed its value at steady state.  Examining \eqref{eq:heat_work_gen}, we see that this describes a piecewise convex function for $\Delta H (Q_1, Q_2)>0$ (convex within each quadrant in $Q_1$ and $Q_2$, centred on the origin), which for constant $\Delta H$ does not permit any real solution for $Q_1$ higher than the steady-state value. % (in other words, the term $X_2^{(a)} |Q_2|^{\alf+1}$ cannot be negative).  
Figure \ref{fig:3D}(b) thus illustrates the {\it simultaneous} action, in parallel pipe systems of constant $\Delta H$, of (i) the MaxEPMod principle of \Zup and co-workers, which discriminates the steady-state position from the set of non-steady-state solutions for each flow regime and specified $\Delta H$, and (ii) a strict inversion of the {Ziegler-Paltridge} MaxEP principle, which provides a method to determine the flow regime at a specified $\Delta H$.

%
%%############################################################################
\section{\label{Concl}Conclusions}
%##############################################
%% 

The present study examines the role of various entropy production extremum principles for the analysis of steady-state flow of an isothermal, incompressible fluid subject to friction in single or parallel pipes.  Firstly, flow in a single pipe is shown to satisfy the {Ziegler-Paltridge} MaxEP principle, which provides a criterion for the choice of laminar or turbulent flow.  The resulting plots of entropy production against Reynolds number (indicating distance from zero-flow equilibrium) are remarkably similar to those reported for other dissipative ``complex systems'' such as heat convection \cite{Schneider_K_1994} and biological metabolism \cite{Bruers_M_2007}, suggesting wide applicability of this principle.  Secondly, flow in two or more dissimilar parallel pipes subject to a constant total flow rate can simultaneously satisfy two principles: (i) a pipe network MinEP principle with respect to the choice of flow rates, and (ii) the {Ziegler-Paltridge} MaxEP principle with respect to the choice of flow regime.  The former principle - different to but allied to that of Prigogine - arises as a simple consequence of the stability of the steady state compared to non-steady-state flows; it is proven for isothermal laminar and turbulent flows in parallel pipes with a constant power law exponent, but is otherwise proven to be invalid.  The latter principle appears to be more fundamental.  Flow in parallel pipes subject to a constant head condition can satisfy: (i) a modified maximum entropy production (MaxEPMod) principle of \Zup and co-workers with respect to the choice of flow rates, and (ii) an inversion of the {Ziegler-Paltridge} MaxEP principle with respect to the flow regime. The former is again valid only for flows with a constant power law exponent. 

In either type of parallel pipe system, the pipe network MinEP or MaxEPMod principles provide the same information as the known closed-form solution and, as shown, are not of universal validity. Such methods may therefore be useful for numerical solution purposes, but cannot be considered as fundamental principles. (Similar considerations apply to the electrical flow analogues of these principles.)  In contrast, the {Ziegler-Paltridge} MaxEP principle or its inversion provides a criterion with which to choose between two known steady-state solutions; it is therefore of much greater importance for the analysis of fluid flow systems. 

Further work is required on the validity of entropy production extremum principles in more general pipe networks, including those with series and parallel flows and with inflows or outflows at each node.

%%############################################################################
%\ack
%\begin{acknowledgments}
\section*{Acknowledgments}
The author is grateful for useful discussions with Bernd Noack, and travel support from the University of Copenhagen, Denmark, the Technische Universit\"at Berlin, Germany, and The University of New South Wales.

%% ############################################################################

%############################################################################
\appendix
%############################################################################
\section{\label{Apx_nonpower} The Pipe Network $\bf{MinEP}$ Principle for Arbitrary $\mathbf{\it f(Re)}$}  

Consider the case of two parallel pipes subject to a constant total flow rate $Q$ (Figure \ref{fig:parallel}), with variable friction factors $f_j(Re_j)$.  From \eqref{eq:DW} and \eqref{eq:EP_network}, with the flow rates $Q_j$ expressed as functions of Reynolds numbers $Re_j$, the entropy production per unit specific weight is:
\begin{equation}
\frac{\dot{\sigma} T}{\rho g} = \frac{\pi \mu^3}{8 \rho^3 g} \Bigl( \frac{f_1 L_1 |Re_1|^3}{d_1^2} + \frac{f_2 L_2 |Re_2|^3}{d_2^2} \Bigr)
\label{eq:EP_case0_nonpower}
\end{equation}
From node continuity \eqref{eq:cty_case0}, $Re_2 = 4 \rho Q / \pi \mu d_2 - Re_1 d_1/d_2$, giving the objective function:
\begin{equation}
F = \frac{\pi \mu^3}{8 \rho^3 g} \Bigl( \frac{ f_1 L_1 |Re_1|^3}{d_1^2} + \frac{f_2 L_2}{d_2^5} \Bigl| \frac{4 \rho Q}{\pi \mu} - {Re_1 d_1} \Bigr|^3 \Bigr)
\label{eq:EP_case_nonpower2}
\end{equation}
Setting the derivative with respect to $Re_1$ to zero, and back-substituting in terms of $Re_2$ and the head losses, gives:
\begin{align}
\begin{split}
\frac {\partial F}{\partial Re_1} =0 &=  \frac{\pi \mu^3}{8 \rho^3 g} \Bigl[ \frac{3 L_1 f_1 Re_1 |Re_1|}{d_1^2} - \frac{3 d_1 L_2 f_2 Re_2 |Re_2|}{d_2^3} 
\\
&+ \frac{L_1 |Re_1|^3}{d_1^2} \frac{\partial f_1}{\partial Re_1} 
- \frac{d_1 L_2 |Re_2|^3 }{d_2^3} \frac{\partial f_2}{\partial Re_2}   \Bigr]
\\
&=  \frac{3 \pi \mu d_1}{4 \rho} (H_{L1}-H_{L2})   
\\
&+ \frac{\pi \mu^3}{8 \rho^3 g} \Bigl[ \frac{L_1 |Re_1|^3}{d_1^2} \frac{\partial f_1}{\partial Re_1} 
\\
&\qquad \qquad - \frac{d_1 L_2 |Re_2|^3}{d_2^3} \frac{\partial f_2}{\partial Re_2}  \Bigr]
\end{split}
\raisetag{20pt}
\label{eq:diff_EP_case_nonpower}
\end{align}
The last form can only satisfy $H_{L1}=H_{L2}$ if it meets the condition:
\begin{equation}
\frac{L_1 |Re_1|^3}{d_1^3} \frac{\partial f_1}{\partial Re_1} 
= \frac{L_2 |Re_2|^3}{d_2^3} \frac{\partial f_2}{\partial Re_2} 
\label{eq:cond_nonpower}
\end{equation}
or, substituting for $L_j$ from \eqref{eq:DW} and the definition of $Re_j$:
\begin{gather}
\frac{H_{L1} Re_1}{f_1} \frac{\partial f_1}{\partial Re_1} 
= \frac{H_{L2} Re_2}{f_2} \frac{\partial f_2}{\partial Re_2} 
\label{eq:cond_nonpower2}
\end{gather}
For $H_{L1}=H_{L2}$, this gives the condition:
\begin{gather}
\frac{Re_j}{f_j} \frac{\partial f_j}{\partial Re_j} = A = \text{constant}
\label{eq:cond_nonpower3}
\end{gather}
Eq.\ \eqref{eq:cond_nonpower3} is a differential equation with general solution:
\begin{gather}
f_j = C {Re_j}^A 
\label{eq:sol_nonpower}
\end{gather}
where $C$=constant. To preserve $f_j \in \mathbb{R}^+$ for $Re_j<0$ and $C>0$ we can, without loss of utility, restrict \eqref{eq:sol_nonpower} to the form:
\begin{gather}
f_j = C |Re_j|^A 
\label{eq:sol_nonpower2}
\end{gather}
which also satisfies \eqref{eq:cond_nonpower3}. This proves that in a parallel pipe network, the MinEP principle conforms to the closed-form solution only for pipes with power-law behaviour \eqref{eq:HL_power} with a common exponent $A=\alf-2$. A similar result ensues for any number $J\ge 2$ of parallel pipes.

\section{\label{Apx_nonpower2} The $\bf{MaxEPMod}$ Principle for Arbitrary $\mathbf{\it f(Re)}$}  

Consider two parallel pipes subject to a constant head (Figure \ref{fig:Zup}(a)), with variable friction factors $f_j(Re_j)$. From \eqref{eq:Zup_Lagr} and the definition of $Re_j$, the objective function is:
\begin{align}
\begin{split}
G &= \frac{(1-\lambda) \pi \mu^3}{8 \rho^3 g} \Bigl( \frac{f_1 L_1|Re_1|^3 }{d_1^2} +  \frac{f_2 L_2 |Re_2|^3 }{d_2^2} \Bigr) 
\\
&+ \frac{\lambda |\Delta H| \pi \mu}{4 \rho} (|Re_1| d_1 + |Re_2| d_2)
\end{split}
\label{eq:Zup_Lagr_gen}
\end{align}
Setting $\partial G/\partial Re_j=0$ and solving for $\lambda$ gives:
\begin{align}
\lambda &= \frac{\mu^2 L_j Re_j^2 \Bigl(3 f_j + Re_j \; \dfrac{d f_j}{d Re_j}  \Bigr)}{\mu^2 L_j Re_j^2 \Bigl(3 f_j + Re_j \; \dfrac{d f_j}{d Re_j} \Bigr)  - 2 d_j^3 \rho^2 g |\Delta H|}
\\
&=\frac{ L_j \Bigl( 3 Re_j^2 f_j + Re_j^3 \; \dfrac{d f_j}{d Re_j} \Bigr) }{L_j \Bigl( 3 Re_j^2 f_j + Re_j^3 \; \dfrac{d f_j}{d Re_j} \Bigr) - 2 |\Delta H| Ga_j}
\\
&=\frac{ 3 + \dfrac{Re_j}{f_j} \dfrac{d f_j}{d Re_j} }{3 + \dfrac{Re_j}{f_j} \dfrac{d f_j}{d Re_j}  - \dfrac{|\Delta H| \sign(Re_j)}{H_{Lj}}}
\end{align}
where all dependent extraneous variables are eliminated by successively substituting for the system properties using the Galileo number (\S\ref{pipeflow}), and for $f_j$ and then $L_j$ in terms of the head loss, $H_{Lj} = \frac{1}{2} f_j L_j Re_j |Re_j| / Ga_j$ (see \eqref{eq:DW}).  For the MaxEPMod principle to be valid, firstly, it must be true that $|\Delta H| \sign(Re_j) =H_{Lj}, \forall j$, and secondly, it must be possible to choose a common $\lambda$ for all pipes. Equating $\lambda$ for two parallel pipes then gives the condition:
\begin{align}
\dfrac{Re_j}{f_j} \dfrac{d f_j}{d Re_j} = A = \text{constant}
\end{align}
This is the same differential equation as \eqref{eq:cond_nonpower3}, again yielding a power-law solution \eqref{eq:sol_nonpower}-\eqref{eq:sol_nonpower2}. The same result is obtained for any number of parallel pipes. This proves that in a parallel pipe network, the MaxEPMod principle is valid only for flow in power-law pipes of common exponent $\alf$.

%###############################
%\section{References*}


\begin{thebibliography}{150}
%% ############################################################################
%EP def
\bibitem{deGroot_M_1962} S.R. de Groot, P. Mazur, Non-Equilibrium Thermodynamics, Dover Publications, NY, 1984.
\bibitem{Prigogine_1967} I. Prigogine, Introduction to Thermodynamics of Irreversible Processes, 3rd ed., Interscience Publ., NY, 1967.
\bibitem{Kondepudi_P_1998} D. Kondepudi, I. Prigogine, Modern Thermodynamics: from Heat Engines to Dissipative Structures, John Wiley \& Sons, Chichester, UK, 1998.
%dissip sys
\bibitem{Prigogine_1980} I. Prigogine, From Being to Becoming: Time and Complexity in the Physical Sciences, W.H. Freeman \& Co., San Francisco, 1980.
\bibitem{Prigogine_S_1984} I. Prigogine, I. Stengers, Order Out of Chaos: Man's New Dialogue with Nature, Fontana Paperbacks, London, 1984.
%fm books
\bibitem{Bird_etal_2006} R.B. Bird, W.E. Stewart, E.N. Lightfoot, Transport Phenomena, 2nd ed., John Wiley \& Sons, NY, 2002.
\bibitem{Munson_etal_2009}B.R. Munson, D.F. Young, T.H. Okiishi, W.W. Huebsch, Fundamentals of Fluid Mechanics, 6th ed., John Wiley, NY, 2009.
%Bejan
\bibitem{Bejan_1982}A. Bejan, Entropy Generation Through Heat and Fluid Flow, John Wiley, NY, 1982.
%\bibitem{Bejan_1984}A. Bejan, Convection Heat Transfer, John Wiley, NY, 1984.
\bibitem{Bejan_1996}A. Bejan, Entropy Generation Minimization, CRC Press, Boca Raton, 1996.
\bibitem{Bejan_2006}A. Bejan, Advanced Engineering Thermodynamics, 3rd ed., John Wiley, NJ, 2006.
%Paulus Jr
\bibitem{PaulusJr_2000}D.M. Paulus Jr, Second Law Applications in Modeling, Design and Optimization, PhD thesis, Marquette University, Milwaukee, WI, 2000 (unpub.).
\bibitem{PaulusJr_G_2004}D.M. Paulus Jr, R.A. Gaggioli,\ti{ Some observations of entropy extrema in fluid flow,} Energy {\vol 29} 2847\ti{-2500} (2004).
\bibitem{Martyushev_2007} L.M. Martyushev,\ti{ Some interesting consequences of the maximum entropy production principle,} J. Exper. Theor. Phys. {\vol 104}(4) 651\ti{-654} (2007).
%thermo
\bibitem{Clausius_1865} R. Clausius,\ti{ \"Uber verschiedene f\"ur die Anwendung bequeme Formen der Hauptgleichungen der mechanischen WŠrmetheorie,} Poggendorfs Annalen {\vol 125}, 335\ti{-400} (1865); English transl.: R.B. Lindsay {\it in} J. Kestin (ed.) {The Second Law of Thermodynamics}, Dowden, Hutchinson \& Ross, PA (1976) 162\ti{-193}.
\bibitem{Boltzmann_1877} L. Boltzmann,\ti{ \"Uber die Beziehung zwischen dem zweiten Hauptsatze dewr mechanischen W\"armetheorie und der Wahrscheinlichkeitsrechnung, respective den S\"atzen \"uber das W\"armegleichgewicht,} Wien. Ber. {\vol 76}, 373\ti{-435} (1877); English transl.: J. Le Roux (2002) 1\ti{-63} {\it http://www.essi.fr/$\sim$leroux/}.
\bibitem{Gibbs_1875} J.W. Gibbs,\ti{ On the equilibrium of heterogeneous substances,} Trans. Connecticut Acad. {\vol 3}, 108\ti{-248} (1875-1876); 343\ti{-524} (1877-1878).
\bibitem{Callen_1985} H.B. Callen, Thermodynamics and an Introduction to Thermostatistics, 2nd ed., John Wiley, NY, 1985.
%critique / review
\bibitem{Martyushev_S_2006} L.M. Martyushev, V.D. Seleznev,\ti{ Maximum entropy production principle in physics, chemistry and biology,} Physics Reports {\vol 426}, 1\ti{-45} (2006).
\bibitem{Martyushev_etal_2007} L.M. Martyushev, A.S. Nazarova, V.D. Seleznev,\ti{ On the problem of the minimum entropy production in the nonequilibrium stationary state,} J. Phys. A: Math. Theor. {\vol 40} 371\ti{-380} (2007).
\bibitem{Bruers_2007c} S. Bruers,\ti{ Classification and discussion of macroscopic entropy production principles,} {\it arXiv:cond-mat/0604482v3}, 2007.  
%max_sigma_diss
\bibitem{Onsager_1931a} L. Onsager,\ti{ Reciprocal relations in irreversible processes I,} Phys. Rev. {\vol 37}, 405\ti{-426} (1931).
\bibitem{Onsager_1931b} L. Onsager,\ti{ Reciprocal relations in irreversible processes II,} Phys. Rev. {\vol 38}, 2265\ti{-2279} (1931).
\bibitem{Onsager_M_1953}L. Onsager, S. Machlup,\ti{ Fluctuations and Irreversible Processes,} Phys. Rev. {\vol 91}(6), 1505\ti{-1515} (1953).
%max_sigma_diss_fm
\bibitem{Helmholtz_1868} H. Helmholtz,\ti{ Zur Theorie der station\"airen Str\"ome in reibenden Fl\"ussigkeiten}, Verh. des naturhist-med. Vereins zu Heidelberg, Bd. v. S. 1-7; Collected Works. {\vol 1}, p. 223.
%\ti{ **,} Wiss. Abh. {\vol 1} 223\ti{-230} (1868). %Verh. Des Maturh. Med. Vereins zu Heidelberg, 5, I***
\bibitem{Rayleigh_1913}Lord Rayleigh,\ti{ On the motion of a viscous fluid,} Phil. Mag. {\vol 26}, 776\ti{-786} (1913).
%\bibitem{Keller_1970}J.B. Keller,\ti{ Extremum principles for irreversible processes,} J. Math. Phys. {\vol 11}(9), 2919\ti{-2931} (1970).
\bibitem{Horne_K_1988}W.C. Horne, K. Karamcheti, Extrema Principles of Entropy Production and Energy Dissipation in Fluid Mechanics, NASA Tech. Memo. 100992, NASA Ames Research Center, CA, 1988.
%max_sigma_diss_elec
\bibitem{Zupanovic_etal_2004}P. \v{Z}upanovi\'c, D. Jureti\'c, S. Botri\'c,\ti{ KirchhoffÕs loop law and the maximum entropy production principle,} Phys. Rev. E {\vol 70}, article 056108 (2004).
\bibitem{Botric_etal_2005}S. Botri\'c, P. \v{Z}upanovi\'c, D. Jureti\'c,\ti{ Is the stationary current distribution in a linear planar electric network determined by the principle of maximum entropy production?,} Croatica Chemica Acta {\vol 78}(2), 181\ti{-184} (2005). 
\bibitem{Christen_2006} T. Christen,\ti{ Application of the maximum entropy production principle to electrical systems,} J. Phys. D: Appl. Phys. {\vol 39}, 4497\ti{-4503} (2006).
\bibitem{Bruers_etal_2007a} S. Bruers, C. Maes, K. Neto\v{c}n\'{y},\ti{ On the validity of entropy production principles for linear electrical circuits,} J. Stat. Phys. {\vol 129}, 725\ti{-740} (2007).
%%%minEP - elec
\bibitem{Jeans_1966}J. Jeans, The Mathematical Theory of Electricity and Magnetism, 5th ed., Cambridge U.P., 1925.
\bibitem{Landauer_1975}R. Landauer,\ti{ Stability and entropy production in electrical circuits,} J. Stat. Phys. {\vol 13}(1), 1-16 (1975).
\bibitem{Jaynes_1980}E.T. Jaynes,\ti{ The minimum entropy production principle,} Ann. Rev. Phys. Chem. {\vol 31}: 579-601 (1980).
%maxEP
\bibitem{Ziegler_1977}H. Ziegler, An Introduction to Thermomechanics, North-Holland Publ. Co., Amsterdam, 1977.
\bibitem{Paltridge_1975}G.W. Paltridge,\ti{ Global dynamics and climate - a system of minimum entropy exchange,} Quart. J. Royal Meteorol. Soc. {\vol 101}, 475\ti{-484} (1975).
\bibitem{Paltridge_1978}G.W. Paltridge,\ti{ The steady-state format of global climate,} Quart. J. Royal Meteorol. Soc. {\vol 104}, 927\ti{-945} (1978).
%MaxEP examples
\bibitem{Ozawa_etal_2001}H. Ozawa, S. Shikokawa, H. Sakuma,\ti{ Thermodynamics of fluid turbulence: A unified approach to the maximum transport properties,} Phys. Rev. E {\vol 64}, article 026303 (2001). 
\bibitem{Kleidon_etal_2003} A. Kleidon, K. Fraedrich, T. Kunz, F. Lunkeit,\ti{ The atmospheric circulation and states of maximum entropy production,} Geophys. Res. Lett. {\vol 30}(23), article 2223 (2003).
\bibitem{Ozawa_etal_2003}H. Ozawa, A. Ohmura, R.D. Lorenz, T. Pujol,\ti{ The second law of thermodynamics and the global climate system: A review of the maximum entropy production principle,} Rev. Geophys. {\vol 41}, article 4 (2003). 
\bibitem{Kleidon_L_book_2005} A. Kleidon, R.D. Lorenz (eds.) Non-equilibrium Thermodynamics and the Production of Entropy: Life, Earth and Beyond, Springer Verlag, Heidelberg, 2005.
\bibitem{Vanyo_Paltridge_1981}J.P. Vanyo, G.W. Paltridge,\ti{ A model for energy dissipation at the mantle-core boundary,} Geophys. J. Royal Astron. Soc. {\vol 66}, 677\ti{-690} (1981).
\bibitem{Kleidon_2004} A. Kleidon,\ti{ Beyond Gaia: thermodynamics of life and Earth system functioning,} Climatic Change {\vol 66}, 271\ti{-319} (2004).
\bibitem{Kleidon_S_2008} A. Kleidon, S. Schymanski,\ti{ Thermodynamics and optimality of the water budget on land: a review,} Geophys. Res. Lett. {\vol 35}, L20404 (2008).
\bibitem{Martyushev_A_2003} L.M. Martyushev, E.G. Axelrod,\ti{ From dendrites and S-shaped growth curves to the maximum entropy production principle,} JETP Letters {\vol 78}(8), 476\ti{-479} (2003).
\bibitem{Christen_2007b} T. Christen,\ti{ Modelling diffusion in nonuniform solids using entropy production rate,} J. Phys. D: Appl. Phys. {\vol 40}, 5723\ti{-5726} (2007).
\bibitem{Meysman_B_2007} F.J.R. Meysman, S. Bruers,\ti{ A thermodynamic perspective on food webs: Quantifying entropy production within detrital-based ecosystems,} J. Theor. Biol. {\vol 249}, 124\ti{-139} (2007).
\bibitem{Bruers_M_2007} S. Bruers, F.J.R. Meysman,\ti{ A useful correspondence between fluid convection and ecosystem operation,} {\it arXiv:0708.0091v1}, 2007.
\bibitem{Juretic_Z_2003} D. Jureti\'c, P. \v{Z}upanovi\'c,\ti{ Photosynthetic models with maximum entropy production in irreversible charge transfer steps,} Computational Biology and Chemistry {\vol 27}, 541\ti{-553} (2003).
\bibitem{Dewar_etal_2006} R.C. Dewar, D. Jureti\'c, P. \v{Z}upanovi\'c,\ti{ The functional design of the rotary enzyme ATP synthase is consistent with maximum entropy production,} Chem. Phys. Lett. {\vol 430}, 177\ti{-182} (2006).
\bibitem{Main_Naylor_2008}I.G. Main, M. Naylor,\ti{ Maximum entropy production and earthquake dynamics,} Geophys. Res. Lett. {\vol 35}: L19311 (2008).
%hydraulics MaxEnt methods
\bibitem{Awumah_etal_1990}K. Awumah, I. Goulter, S.K. Bhatt,\ti{ Assessment of reliability in water distribution using entropy based measures,} Stochastic Hydrol. Hydraul. {\vol 4} 309\ti{-320} (1990). 
\bibitem{Tanyimboh_T_1993}T.T. Tanyimboh, A.B. Templeman,\ti{ Calculating maximum entropy flows in networks,} J. Oper. Res. Soc. {\vol 44}(4) 383\ti{-396} (1993) 
\bibitem{deSchaetzen_etal_2000}W.B.F. de Schaetzen, G.A. Walters, D.A. Savic,\ti{ Optimal sampling design for model calibration using shortest path, genetic and entropy algorithms,} Urban Water {\vol 2} 141\ti{-152} (2000).
\bibitem{Formiga_etal_2003} L.T.M. Formiga, F.H. Chaudhry, P.B. Cheung, L.F.R. Reis,\ti{ Optimal design of water distribution system by multiobjective evolutionary methods,} in C.M. Forseca (eds), EMO 2003, LNCS 2632, 677\ti{-691} (2003). 
\bibitem{Ang_J_2003}W.-K. Ang, P.W. Jowitt,\ti{ Some observations on energy loss and network entropy in water distribution networks,} Eng. Opt. {\vol 35}(4) 375\ti{-389} (2003).
\bibitem{Setiadi_etal_2005}Y. Setiadi, T.T. Tanyimboh, A.B. Templeman,\ti{ Modelling errors, entropy and the hydraulic reliability of water distribution systems,} Adv. Eng. Software {\vol 36} 780\ti{-788} (2005).
%my MEP
\bibitem{Niven_MEP}R.K. Niven,\ti{ Steady state of a dissipative flow-controlled system and the maximum entropy production principle}, Phys. Rev. E {\vol 80}(2): 021113 (2009); also http://arxiv.org/abs/0902.1568.
\bibitem{Niven_MaxEnt09}R.K. Niven,\ti{ Jaynes' MaxEnt, steady state flow systems and the maximum entropy production principle,} in P. Goggans (ed.), MaxEnt09, Oxford, MS, 5-10 July 2009, AIP, in press (http://arxiv.org/abs/0908.0990).
\bibitem{Niven_PhilTransB}R.K. Niven,\ti{ Minimisation of a free-energy-like potential for non-equilibrium systems at steady state,} in submission to Phil. Trans. B, http://arxiv.org/abs/0908.2268v1.
%MaxEP derivs
\bibitem{Dewar_2003}R.C. Dewar,\ti{ Information theory explanation of the fluctuation theorem, maximum entropy production and self-organized criticality in non-equilibrium stationary states,} J. Phys. A: Math. Gen. {\vol 36}, 631\ti{-641} (2003).
\bibitem{Dewar_2005}R.C. Dewar,\ti{ Maximum entropy production and the fluctuation theorem,} J. Phys. A: Math. Gen. {\vol 38}, L371\ti{-L381} (2005).
\bibitem{Attard_2006a} P. Attard,\ti{ Statistical mechanical theory for steady state systems. VI. Variational principles,} J. Chem. Phys. {\vol 125}, article 214502 (2006).
\bibitem{Attard_2006b} P. Attard,\ti{ Theory for non-equilibrium statistical mechanics,} Phys. Chem. Chem. Phys. {\vol 8}, 3585\ti{-3611} (2006).
\bibitem{Zupanovic_etal_2006}P. \v{Z}upanovi\'c, S. Botri\'c, D. Jureti\'c,\ti{ Relaxation processes, MaxEnt formalism and Einstein's formula for the probability of fluctuations,} Croatica Chemica Acta {\vol 79}(3), 335\ti{-338} (2006). 
\bibitem{Martyushev_2007b} L.M. Martyushev,\ti{ Do nonequilibrium processes have features in common?,}  http://arxiv.org/abs/0709.0152 (2007).
%constructal
\bibitem{Bejan_Lorente_2006}A. Bejan, S. Lorente,\ti{ Constructal theory of generation of configuration in nature and engineering,} J. Appl. Phys. {\vol 100}, 041301 (2006).
\bibitem{Bejan_2007}A. Bejan,\ti{ Constructal theory of pattern formation,} Hydrol. Earth Syst. Sci. {\vol 11} 753\ti{-768} (2007). 
%Salamon bros
\bibitem{Salamon_A_G_B_1980}P. Salamon, B. Andresen, P.D. Gait, R.S. Berry,\ti{ The significance of Weinhold's length,} J. Chem. Phys. {\vol 73}(2) 1001-1002 (1980), erratum {\vol 73}(10) 5407 (1980).
\bibitem{Salamon_B_1983}P. Salamon, R.S. Berry,\ti{ Thermodynamic length and dissipated availability,} Phys. Rev. Lett. {\vol 51}(13) 1127-1130 (1983).
\bibitem{Nulton_etal_1985}J. Nulton, P. Salamon, B. Andresen, Q. Anmin,\ti{ Quasistatic processes as step equilibrations,} J. Chem. Phys. {\vol 83}(1) 334-338 (1985).
\bibitem{Andresen_G_1994}B. Andresen, J.M. Gordon,\ti{ Constant thermodynamic speed for minimizing entropy production in thermodynamic processes and simulated annealing,} Phys. Rev. E {\vol 50}(6) 4346\ti{-4351} (1994).
\bibitem{Crooks_2007}G.E. Crooks,\ti{ Measuring thermodynamic length,} Phys. Rev. Lett. {\vol 99} 100602 (2007).
%my Riemannian
\bibitem{Niven_A_2009}R.K. Niven, B. Andresen,\ti{ Jaynes' maximum entropy principle, Riemannian metrics and generalised least action bound,} in Dewar, R.L., Detering, F. (eds) Complex Physical, Biophysical and Econophysical Systems, World Scientific Lecture Notes in Complex Systems, Vol. 9, %Proc. 22nd Canberra International Physics Summer School, The Australian National University, Canberra, 8-19 December 2008, 
World Scientific, Hackensack, NJ, http://arxiv.org/abs/0907.2732.
%
%fluid mech
\bibitem{Street_etal_1996} R.L. Street, G.Z. Watters, J.K. Vennard (1996) Elementary Fluid Mechanics, 7th ed., John Wiley, NY. 
\bibitem{Schlichting_2001} H. Schlichting, K. Gersten, {\it Boundary Layer Theory}, 8th ed., Springer, NY, 2001.
%Galileo no
\bibitem{Pavlov_etal_1979}K.F. Pavlov, P.G. Romankov, A.A. Noskov, Examples and Problems to the Course of Unit Operations of Chemical Engineering, English transl., Mir Publishers, Moscow, 1979.
\bibitem{Niven_2002}R.K. Niven,\ti{ Physical insight into the Ergun and Wen \& Yu equations for fluid flow in packed and fluidised beds,} Chem. Eng. Sci. {\vol 57}, 527\ti{-534} (2002).
%elec systems
\bibitem{Stephen_1960} K. Stephen, Electrical Circuit Analysis, Macmillan, 1960.
\bibitem{Morris_1993} N.M. Morris, Electrical Circuit Analysis and Design, Palgrave Macmillan, 1993.
%convection
\bibitem{Schneider_K_1994}E.D. Schneider, J.J. Kay,\ti{ Life as a manifestation of the second law of thermodynamics,} Math. Comput. Modelling {\vol 19}(6-8) 25\ti{-48} (1994).
%Ergun
\bibitem{Niven_2002}R.K. Niven, \ti{ Physical insight into the Ergun and Wen \& Yu equations for fluid flow in packed and fluidised beds,} Chem. Eng. Sci. {\vol 57} 527\ti{-534} (2002).
%osc pipe etc
\bibitem{Benhamou_etal_2004} B. Benhamou, A. Laneville, N. Galanis,\ti{ Transition to turbulence: The case of a pipe in radial oscillations,} Int. J. Thermal Sci. {\vol 43} 1141\ti{-1151} (2004). 
\bibitem{Li_etal_2009}C.W. Li, L. Jia, T.T. Zhang,\ti{ The entrance effect on gases flow characteristics in micro-tube,} J. Thermal Sci. {\vol 18}(4), 353\ti{-357} (2009).

\end{thebibliography}
\end{document}